\documentclass[12pt,letterpaper]{article}
\usepackage{jheppub}
\usepackage{slashed}
\usepackage{amsmath}
\usepackage{amssymb}
\usepackage{epsfig}

\def\({\left(} \def\){\right)}
\def\[{\left[} \def\]{\right]}
\def\del{{\partial}}
\newcommand{\non}{\nonumber \\}

\newcommand{\be}{\begin{equation}}
\newcommand{\ee}{\end{equation}}
\newcommand{\bea}{\begin{eqnarray}}
\newcommand{\eea}{\end{eqnarray}}
\newcommand{\ba}{\begin{eqnarray}}
\newcommand{\ea}{\end{eqnarray}}

\newcommand{\beq}{\begin{equation}}
\newcommand{\eeq}{\end{equation}}
\newcommand{\beqa}{\begin{eqnarray}}
\newcommand{\eeqa}{\end{eqnarray}}
\newcommand{\beqar}{\begin{eqnarray*}}
\newcommand{\eeqar}{\end{eqnarray*}}

\newcommand{\reef}[1]{(\ref{#1})}
\newcommand{\ssc}{\scriptscriptstyle}
\newcommand{\eg}{{\it e.g.,}\ }
\newcommand{\ie}{{\it i.e.,}\ }

\renewcommand{\a}[1]{a_\mt{#1}}

\newcommand{\mt}[1]{\textrm{\tiny #1}}

\newcommand{\gas}{\gamma_{\ssc D}}

\newcommand{\hi}{{\hat \imath}}

\newcommand{\cA}{{\cal A}}

\newcommand{\m}{v}

\newcommand{\cR}{{\mathcal R}}

\newcommand{\cM}{{\mathcal M}}

\newcommand{\ren}{R\'enyi\ }

\newcommand{\tr}{{\rm tr}}
\newcommand{\Tr}{\text{Tr}}

\def\a{\alpha}

\catcode`\@=12

\newcommand{\BT}{\mathbb{T}}
\newcommand{\BI}{\mathbb{I}}
\newcommand{\BS}{\mathbb{S}}
\newcommand{\BR}{\mathbb{R}}
\newcommand{\BH}{\mathbb{H}}

\newcommand{\al}{\alpha}

\newcommand{\te}{t_\mt{E}}

\def\del          {\partial}

\def\tr           {\mathop{\rm Tr}}

\def\call         {{\cal L}}
\def\calm         {{\cal M}}
\def\caln         {{\cal N}}
\def\calo         {{\cal O}}

\def\calv         {{\cal V}}

\renewcommand{\eqref}[1]{(\ref{#1})}

\begin{document}

\preprint{arXiv:1210.nnnn [hep-th]}

\title{Observations on entanglement entropy\\
in massive QFT's}

\author{Aitor Lewkowycz,$^{1,2}$ Robert C. Myers$^{1}$ and Michael Smolkin$^{1}$\\

\vskip .5cm

{\it $^1$Perimeter Institute for Theoretical Physics,
\\ \ \,31 Caroline Street North, Waterloo,
Ontario N2L 2Y5, Canada\\
$^2$Department of Physics, Princeton University, Princeton, NJ 08544, USA}}

\vskip .5cm

\abstract{We identify various universal contributions to the entanglement
entropy for massive free fields. As well as the `area' terms found in
\cite{wilczek}, we find other geometric contributions of the form discussed in
\cite{relevant}. We also compute analogous contributions for a strongly coupled
field theory using the AdS/CFT correspondence. In this case, we find the
results for strong and weak coupling do not agree.}

\maketitle

\section{Introduction}

Entanglement entropy has emerged as a topic of interest in a wide
variety range of research areas ranging from condensed matter physics
\cite{wenx} to quantum gravity \cite{bh9,mvr}. In the context of
quantum field theory (QFT), when one considers the entanglement between
two regions,\footnote{These are spatial regions on a fixed Cauchy
surface.} one finds that the entanglement entropy is UV divergent
because of short range correlations in the vicinity of the `entangling
surface' $\Sigma$ separating the two regions. If the calculation is
regulated with a short distance cut-off $\delta$, the leading
contributions for a QFT in $D$ spacetime dimensions generically take
the form
 \be
 S_\mt{EE}=\frac{c_2}{\delta^{D-2}}+\frac{c_4}{\delta^{D-4}}
 +\frac{c_6}{\delta^{D-6}}+\cdots\,,
 \label{diverg0}
 \ee
where each of the coefficients $c_{2k}$ involves an integration over the
boundary $\Sigma$. For example, the leading term then yields the famous `area
law' result with $c_2\propto\cA_\Sigma$ \cite{bh9}. Unfortunately the
coefficients appearing in these power law divergent terms above are sensitive
to the details of the UV regulator --- for further discussion, see section
\ref{discuss}.

However, certain subleading contributions can reveal universal data
describing the character and/or the state of the underlying QFT. A
well-known example of such a contribution arises for conformal field
theories (CFT's) in an even number of spacetime dimensions
\cite{finn,adam,rt2,solo,cthem2,EEGB,ben9}. Here, in calculating the
entanglement entropy, one typically finds a logarthmic contribution
$\log(\delta)$ where the coefficient is some linear combination of the
central charges appearing in the trace anomaly of the CFT. The precise
linear combination is again determined by an integral of various
geometric factors over the entangling surface. For a four-dimensional
CFT, this universal contribution takes the form \cite{solo}
 \be
S_\mt{univ}= \frac{\log(\delta/L)}{2\pi} \int_{\Sigma}
d^{2}\sigma\sqrt{h}\,\left[\, a\, \cR_\Sigma -\,c\,
\left(C^{abcd}\,h_{ac} \,h_{bd} - K^{\hi}_a{}^b K^{\hi}_b{}^a+{1\over
2} K^{\hi}_a{}^a K^{\hi}_b{}^{\,b}\right)\,\right]\,,
 \label{WaldS9}
 \ee
where $a$ and $c$ are the usual central charges which appear in the trace
anomaly \cite{traca}. The various geometric factors include: $h_{ab}$,
$\cR_\Sigma$, $K^{\hi}_{ab}$, respectively, the induced metric,
intrinsic scalar curvature and extrinsic curvature of $\Sigma$;
$C_{abcd}$, the (pull-back of the) Weyl curvature of the background
geometry; and $L$, some characteristic scale in the geometry.

A similar class of universal contributions were identified in
\cite{wilczek} for massive QFT's.\footnote{Related results were found
previously in \cite{furwave} and \cite{reallyold}.} In particular,
considering free massive scalar fields, the following universal
contribution to the entanglement entropy was found
\begin{equation}
S_\mt{univ}= \left\lbrace \begin{matrix}
  \ \gas\ \cA_{\Sigma}\ m^{D-2}\,\log(m\delta) &
      \qquad {\rm for\ even}\ D\,,   \\
   \ \gas\ \cA_{\Sigma}\ m^{D-2}\hfill  &
      \qquad {\rm for\ odd}\ D\,.\hfill
\end{matrix} \right.
\label{franks}
\end{equation}
where $m$ is the mass of the scalar and $\cA_{\Sigma}$ is the area of
the entangling surface. For the free scalar theory, the numerical
coefficient $\gas$ is given by \cite{wilczek}
 \be
\gamma_{{\ssc D},\,scalar}= \left\lbrace \begin{matrix}
  \ \frac{(-)^{D/2}}{6\,(4\pi)^{\frac{D-2}{2}}\,
\Gamma(D/2)} &
      \qquad {\rm for\ even}\ D\,,   \\[1em]
   \ \frac{(-)^{\frac{D-1}{2}}\,\pi}{
12\,(4\pi)^{\frac{D-2}{2}}\,\Gamma(D/2)}   &
      \qquad {\rm for\ odd}\ D\,.\hfill
\end{matrix} \right.
 \label{coeff}
 \ee

Motivated by these free field results, ref.~\cite{relevant} began a study of
analogous terms for strongly coupled field theories using holographic
techniques \cite{rt2,rt1}. This approach allows one to study the effect of
perturbing a UV fixed point QFT by introducing general relevant operators,
beyond simple mass terms. The results reveal a broad class of new universal
contributions to entanglement entropy, which can be schematically represented
as
 \be
S_\mt{univ} = \gamma(D,n) \int_{\Sigma}\!d^{D-2}\!\sigma\, \sqrt{h}\
\left[\,``curvature"\,\right]^n\,\times\, \left\lbrace \begin{matrix} \
m^{D-2-2n}\,\log m\delta
& \quad {\rm for\ even}\ D\,,   \\[1em]
\ m^{D-2-2n} \hfill & \quad {\rm for\ odd}\ D\hfill  \,,
\end{matrix} \right.
 \label{univEE}
 \ee
where $m$ is the mass scale appearing in the coupling of the relevant operator.
The schematic geometric factor denotes a combination of both the background and
extrinsic curvatures with a combined dimension $2n$. Note that $0\le n\le
(D-2)/2$ in these expressions. Hence in general, the entanglement entropy
contains a family of universal contributions that includes the `area' terms in
eq.~\reef{franks} at $n=0$ and purely geometric terms, analogous to those in
eq.~\reef{WaldS9}, at $n=(D-2)/2$. Ref.~\cite{relevant} focussed on universal
contributions proportional to $\log(\delta)$, however, note that in general
these terms \reef{univEE} can appear for either even or odd $D$ if the operator
dimension of the relevant deformation is chosen appropriately.\footnote{For
example, choosing an operator with $\Delta=(D+2)/2$, the corresponding coupling
would take the form $g\, m^{(D-2)/2}$ where $g$ is a dimensionless coefficient.
In this case, an `area' term with $n=0$ appears for all $D\ge3$ with
$\gamma\propto g^2$.} The analysis of \cite{relevant} also readily extends to
show the appearance of new universal terms generalizing the cut-off independent
terms in eq.~\reef{franks} for odd $D$. An important distinction is, however,
that generally these cut-off independent terms depend on the underlying state
of the boundary theory while the logarithmic terms are state independent.

One of our objectives in the following is to extend the analysis of
\cite{wilczek} to reveal the new curvature terms appearing in
eq.~\reef{univEE}. This is easily accomplished by performing the free field
calculations on a curved background. In particular, ref.~\cite{wilczek} work
with a `waveguide' geometry, $\BR^{2}\times \BI^{D-2}$, where $\BI$ is a finite
interval with either Dirichlet or Neumann boundary conditions imposed at the
endpoints. In section \ref{wave}, we perform analogous calculations on a
`spherical waveguide', $\BR^{2}\times \BS^{D-2}$. Our analysis also provides a
number of other interesting extensions of that in \cite{wilczek}. As well as
considering the entanglement entropy, we also calculate the \ren entropy, which
is another useful measure of entanglement \cite{renyi0,karol}. In section
\ref{scalar}, we consider a free massive scalar field but we also include the
possibility of the curvature coupling $\frac12 \xi R\, \phi^2$. In section
\ref{fermion}, we extend the analysis to consider a free massive fermion. For
the fermions, we find that the coefficients of the universal `area' terms in
eq.~\reef{franks} become
 \be
\gamma_{{\ssc D},\,fermion}= 2^{ \left\lfloor\! {D\over
2}\!\right\rfloor-1} \ \gamma_{{\ssc D},\,scalar}=\left\lbrace
\begin{matrix}
  \ \frac{(-)^{D/2}}{6\,(2\pi)^{\frac{D-2}{2}}\,
\Gamma(D/2)} &
      \qquad {\rm for\ even}\ D\,,   \\[1em]
   \ \frac{(-)^{\frac{D-1}{2}}\,\pi}{
12\sqrt{2}\,(2\pi)^{\frac{D-2}{2}}\,\Gamma(D/2)}   &
      \qquad {\rm for\ odd}\ D\,,\hfill
\end{matrix} \right.
 \label{coeff2}
 \ee
as was noted previously in \cite{new}. Above, $\lfloor D/2\rfloor$ denotes the
integer part of $D/2$. Our analysis in these sections also allows us to
identify a particular curvature contribution \reef{univEE} to the entanglement
entropy as
\begin{equation}
S_\mt{univ}= \hat\gamma_{D}\,\int_\Sigma
\!d^{D-2}\!\sigma\,\sqrt{h}\ \cR(h)\times \left\lbrace
\begin{matrix}
  \ m^{D-4}\,\log(m\delta) &
      \qquad {\rm for\ even}\ D\ge4\,,   \\
   \ m^{D-4}\hfill  &
      \qquad {\rm for\ odd}\ D\ge5\,.\hfill
\end{matrix} \right.
\label{newX}
\end{equation}
where $\cR(h)$ is the Ricci scalar of the metric induced on the entangling
surface. For a free scalar with a curvature coupling, the new numerical
coefficient $\hat\gamma_{D}$ is given by
 \be
\hat\gamma_{{\ssc D},\,scalar}= \frac{D-2}{2}\,\left(\xi -\frac16 \right)\
\gamma_{D,scalar}
 \label{coeff9x}
 \ee
where $\gamma_{D,scalar}$ is precisely the coefficient appearing in
eq.~\reef{coeff}. For a massive free fermion, this coefficient can be expressed
as
 \be
\hat\gamma_{{\ssc D},\,fermion}= \frac{D-2}{24}\ \gamma_{D,fermion}
 \label{coeff8x}
 \ee
where $\gamma_{D,fermion}$ is given by eq.~\reef{coeff2}.

In section \ref{strong}, we turn to a holographic calculation of entanglement
entropy for the strongly coupled $\caln=2^*$ gauge theory, a massive
deformation of the celebrated $\caln=4$ super-Yang-Mills theory \cite{pw,bpp}.
While similar calculations already appear in \cite{relevant}, the details of
the boundary mass terms and their translation to the dual gravity theory are
precisely understood in this well-studied framework \cite{n22,n23,opera}.
Hence, we are able to compare the `area' contribution \reef{franks} at strong
coupling from holography to the weak coupling results, which combine
eqs.~\reef{coeff} and \reef{coeff2}. Our final conclusion is that the strong
coupling result does {\it not} match the corresponding contribution
\reef{franks} found at weak coupling!

We conclude the paper with a brief discussion of our results and future
directions in section \ref{discuss}. This is followed by two appendices
presenting various technical results: Appendix \ref{appx:sepvar} explicitly
demonstrates the validity of the separation of variables \reef{sepheat} used in
section \ref{fermion} for spin-${1\over 2}$ fields. Finally, Appendix
\ref{renyiSd} repeats the calculations for section \ref{wave} for a
`hyperbolic' waveguide $\BR^2\times\BH^{D-2}$.

\section{\ren entropy on a spherical waveguide} \label{wave}

In general to define the entanglement or \ren entropies of some quantum
system, we begin by dividing the degrees of freedom into two subsets,
$A$ and $\bar A$, and calculate the reduced density matrix $\rho_A={\rm
Tr}_{\bar A}\left(\rho\right)$ where $\rho$ describes the global state
of the system. In a QFT, this is typically realized by beginning with a
Cauchy surface in a fixed background and introducing of an `entangling
surface' $\Sigma$, which divides this surface in two separate regions,
again, denoted $A$ and $\bar A$. The reduced density matrix $\rho_A$ is
then given by integrating over all field configurations in the region
$\bar A$. Given this density matrix, the entanglement entropy is then
defined by the standard von Neumann formula
 \be
S_\mt{EE} = -{\rm Tr}\left(\rho_A\, \log\rho_A\right)\,.
 \label{defEE}
 \ee
while the \ren entropy is given by \cite{renyi0}
 \be
S_\alpha = \frac{1}{1-\alpha}\ \log\,{\rm Tr}\left(\rho_A^{\
\alpha}\right)\,.
 \label{defREN}
 \ee
The latter is usually evaluated for (positive) integer values of $\al$,
in which case, eq.~\reef{defREN} involves a somewhat simpler
calculation since it does not require evaluating the logarithm of
$\rho_A$ appearing in eq.~\reef{defEE}. Further, if the result of
$S_\alpha$ can be continued to real values of $\alpha$, (as will be
possible in the following,) the entanglement entropy can be recovered
as the limit: $S_\mt{EE}=\lim_{\al\to1} S_\al$.

The trace required in eq.~\reef{defREN} has a standard path integral
representation, \eg \cite{cardy0,cardy1},
 \be
\tr\left(\rho_{A}^{\ \alpha}\right)=Z_\alpha/(Z_1)^\al\,,
 \label{replica1}
 \ee
where $\al$ is again assumed to take integer values for the moment.
Implicitly, the first step here is to Wick rotate the background
geometry to Euclidean time $\te=it$. Then $Z_\al$ is the partition
function of the QFT evaluated on an $\al$-fold cover of the Euclidean
background where a cut is introduced throughout region $A$ on the
Cauchy surface, which we denote $\te=0$. At the cut, the fields on the
$k$'th sheet are joined to the fields on the ($k$+1)'th sheet when
approaching from $\te\to0^-$ and to those on the ($k$--1)'th sheet when
approaching from $\te\to0^+$. Hence $Z_\al$ is the partition function
evaluated on a singular covering geometry with an angular excess of
$2\pi(\al-1)$ at the entangling surface $\Sigma$. The factors of the
standard partition function $Z_1$ appear in the denominator of
eq.~\reef{replica1} to ensure that the density matrix is properly
normalized with $Tr\left(\rho_{A}\right)=1$. Given these partition
functions, the \ren entropy \reef{defREN} becomes
 \be
 S_\al={\log Z_\al - \al \log Z_{1}\over 1-\al}\,.
 \label{renyi}
 \ee

Before proceeding with our explicit calculations, let us introduce the
following shorthand notation for simplicity
 \be
 d\equiv D-2~. \label{dD}
 \ee

Now following \cite{wilczek}, our calculations here will focus on free
fields in a waveguide geometry. In particular, throughout this section,
the background geometry will take the form $\cM=\BR^{2}\times \BS^{d}$,
where $\BR^2$ is covered by Cartesian coordinates, $\te$ and $x$. As in
the above discussion, the Cauchy surface is selected by simply setting
$\te=0$ in which case the resulting spatial slice has the waveguide geometry
$\BR\times \BS^{d}$. This slice is then divided into two halves by
choosing the entangling surface to be the sphere at $x=0$. Implicitly
we assume that the QFT on $\cM$ is in its ground state and as described
above, we are considering the reduced density matrix on the region
$A=\lbrace \te=0,\, x>0\rbrace$ resulting after integrating out the
field degrees of freedom in $\bar A$.

To calculate the \ren entropy as described above, we need to evaluate
the partition function on $\cM_\al$, the $\al$-fold cover of $\cM$. At
this point, we note that the background geometry has a rotational
symmetry in the plane around the point $(\te,x)=(0,0)$, which serves as
our entangling surface. Hence in constructing $\cM_\al$, we are simply
replacing the $\BR^2$ component of $\cM$ by a two-dimensional cone
$C_\al$ with an angular excess of $2\pi(\alpha-1)$ at the origin, \ie
$\cM_{\al} = C_{\al} \times \BS^{d}$. For later calculations, we
explicitly write the metric on $C_\al$ as\footnote{After changing
variables $r\rightarrow r/\alpha$, $\theta\rightarrow \alpha\theta$,
one can readily conclude that the angular excess is given by
$2\pi(\alpha-1)$. However, we use the present coordinates \reef{cone}
in the following, since the angular momentum operator takes the
standard form in this representation of the cone.}
 \be
 ds^2=\alpha^{-2}(dr)^2+r^2(d\theta)^2~,
 \label{cone}
 \ee
where $r$ and $\theta$ possessing the full radial and angular range
$0\leq r \leq\infty$, $0\leq \theta\leq 2\pi$. Now this geometry has no
distinguishing features which prefer integer values of $\al$ (apart
from $\al=1$) and so from this point forward, we allow $\al$ to take
any (positive) real value. That is, we are analytically continuing
$\alpha$ already in the covering geometry \cite{wilczek2} rather than
first evaluating $Z_\al$ for integer $\al$ and then analytically
continuing. The rotational symmetry in the transverse space around the
entangling surface is an essential ingredient for this geometric
approach.\footnote{See \cite{cthem2} for further discussion.}

One feature, which distinguishes our background here from that in
\cite{wilczek}, is that the cross-section of the waveguide $\BS^d$ is
curved and hence our calculations of the entanglement entropy below can
reveal new universal contributions of the form given in
eq.~\reef{univEE}. After a few more preliminary remarks, we will
consider a massive free scalar field in section \ref{scalar}, with the
action
 \be
S(\phi)= \int_{\cM_\al}\!\!d^Dx\,\sqrt{g}\ {1\over 2}
\Big(\left(\nabla\phi\right)^2+m^2\phi^2+\xi \, \cR\, \phi^2\Big)~,
 \label{scalaract}
 \ee
where $\cR$ is the Ricci scalar of the background geometry. Hence we
have included a non-minimal coupling to the curvature of the background
 here. Of course, with $m=0$ and $\xi={D-2\over 4(D-1)}$, we
have a conformal scalar field theory. In section \ref{fermion}, we also
consider a massive free fermion field (with minimal coupling to the
background geometry), which becomes a conformal theory if $m=0$.

For either of the above classes of theories, the partition function is
Gaussian and can be exactly evaluated using the heat kernel approach,
\eg \cite{vass}
 \be
\log Z_\al^{(s)}={e^{i2\pi s}\over 2} \int_{\delta^2}^\infty {dt \over
t} \ \Tr\,K^{(s)}_{\cM_\al}\ e^{-t m_{s}^2}~,
 \label{spartition}
 \ee
where $K^{(s)}_{\cM_\al}(t,x,y)$ with $x,y\in\cM_\al$ is the heat
kernel of the corresponding {\it massless} wave operator on $\cM_\al$.
The trace of the heat kernel involves taking the limit of coincident
points, \ie $y\to x$, and integrating over the remaining position $x$.
Of course, a trace is also taken over the spinor indices in the case of
the spin-$1\over 2$ field --- see below. In the above expression and
throughout the following, we use $s= 0$ or ${1\over 2}$ to indicate the
scalar or fermion cases, respectively. Further, $\delta$ is a
short-distance scale introduced to regulate any potential UV
divergences (as discussed in the introduction). Finally $m_{s}$ denotes
the `effective' mass of the field under study. For the fermion, we have
simply $m_{s={1\over 2}}=m$, however, given the non-minimal coupling of
the scalar in eq.~\reef{scalaract}, we have
 \be
 m_{s=0}^2=m^2+\xi\,{d(d-1)\over R^2}~, \label{mass0}
 \ee
where the second term comes from the curvature of the $\BS^{d}$, \ie
$\cR(\BS^{d})=d(d-1)/R^2$ for a sphere of radius $R$.

Notice that the curvature of the full background geometry contains a
singularity at the tip of the cone, \eg
 \be
\cR(\cM_\al)=\cR(\BS^{d})+\cR(C_\al) =  {d(d-1)\over R^2}+
4\pi(1-\al)\,\delta^{(2)}(\vec r) +\ldots
 \label{ricci}
 \ee
where $\vec r\in C_\al$ and the second term corresponds to the leading
contribution from $\cR(C_\al)$ in an expansion near $\al\simeq1$. To
treat this singularity in a well defined way, we delete the point at
the tip of $C_\al$ and work on the space $(C_\al-\{0\})\times\BS^{d}$.
Of course, this means that appropriate boundary conditions must be
imposed at the tip to make the wave operator self-adjoint. This then
becomes the requirement that we must use only non-singular
eigenfunctions when the heat kernel is constructed \cite{kabat}. To
simplify the notation we use the notation $\cM_\al=C_\al\times\BS^d$ to
denote the punctured manifold in what follows.

The wave operators are separable on the product manifold
$C_\al\times\BS^d$ and hence the heat kernel on $\cM_\al$ can be
expressed as the product of the two individual heat kernels on $C_\al$
and $\BS^{d}$, \ie
 \be
 K^{(s)}_{\cM_\al}=\ K_{C_\al}^{(s)}\,K_{\BS^{d}}^{(s)}\,,
 \label{sepheat}
 \ee
where for brevity we have suppressed the arguments of the heat kernels
here. Note that while this separation of variables is obvious in the
case of the spin-$0$ field, it is less evident in the case of the
spin-${1\over 2}$ field due to spinor structure of the heat kernel.
Hence we show that separation of variables indeed holds in the latter
case in Appendix \ref{appx:sepvar}. Given eq.~\reef{sepheat}, one can
write
 \be
\text{Tr}\,K^{(s)}_{\cM_\al}= \text{Tr}\,K_{C_\al}^{(s)}\
\text{Tr}\,K_{\BS^{d}}^{(s)}\,,
 \label{traces}
 \ee
where each trace on the right-hand side involves an integration over the
corresponding component of the product manifold. In the case of spin-${1\over
2}$ field, there is also trace over spinor indices. Using the conventions
established in Appendix \ref{appx:sepvar}, we can regard the two traces on the
right as also including a separate trace over the spinor spaces of the two
component manifolds, $C_\al$ and $\BS^d$.

In eq.~\reef{spartition}, the possible UV divergences at $t\to0$ in the
partition function were regulated by introducing the short-distance
cut-off $\delta$. However, we would also like to introduce a
$\zeta$-function regularization \cite{vass} here since it readily
allows us to identify the universal contributions to the \ren entropy
for general values of $d$. This approach will be applied in calculating
to the \ren entropy of the scalar field on $C_\al\times\BS^d$ in the
next subsection. However, we also apply this regularization to produce
general results for both scalars and fermions on the hyperbolic
waveguide $C_\al\times\BH^d$ in Appendix \ref{renyiSd}.

In the $\zeta$-function approach, the partition function is regulated
by shifting the power of $t$ in eq.~\reef{spartition}
 \be
\log Z_\al^{(s)}={e^{i 2\pi s}\over 2}\, \delta^{-2z} \int_{0}^\infty
{dt \over t^{1-z}} \, \text{Tr}\,K^{(s)}_{\cM_\al} \, e^{-t m_{s}^2}~,
  \label{spartition1}
 \ee
where now $\delta$ appears to keep the whole expression dimensionless.
Of course, after carrying out the integral over $t$, the regulator must
be removed by taking the limit $z\rightarrow 0$ and suitably
renormalizing the parameters of the theory to eliminate possible
divergences in $z$.

Now, as shown in \cite{kabat,jackiw}, the trace of the heat kernel on a cone
depends on $\al$ only, therefore substituting eq.~\reef{traces} into
eq.~\reef{spartition1} yields
 \be
\log Z_\al^{(s)}={e^{i 2\pi s}\over 2} \,\delta^{-2z} \
\text{Tr}K_{C_\al}^{(s)} \ \Gamma(z)\,\zeta_{\BS^d}^{(s)}(z)
 \label{spartition2}
 ~,
 \ee
where the $\zeta$-function is defined as follows
 \be
\zeta^{(s)}_{\BS^d}(z)={1\over \Gamma(z)}\int_0^{\infty}dt \, t^{z-1}
\, \text{Tr} \, K_{\BS^{d}}^{(s)} \, e^{-tm_{s}^2}~.
 \label{zetafunc}
 \ee
Expanding  eq.~\reef{spartition2} in the vicinity of $z=0$ then yields
 \be
\log Z_\al^{(s)}={e^{i 2\pi s}\over 2} \ \text{Tr}K_{C_\al}^{(s)}\
\[\, {\zeta_{\BS^d}^{(s)}(0) \over z} + \left.{ d\zeta_{\BS^d}^{(s)}
\over dz} \right|_{z=0}-\zeta_{\BS^d}^{(s)}(0)\log \delta^2 + \mathcal
O (z)
\] ~,
 \label{expandz}
 \ee
where we rescaled $\delta^2 \rightarrow e^{-\gamma} \delta^2$ to absorb
a term proportional to the Euler constant $\gamma$. The pole term in
the above expression must be removed by suitably renormalizing the
field theory parameters.\footnote{Appearance of the logarithmic term in
eqs.~\reef{expandz} and \reef{spartition3} might seem misleading. Indeed, such terms are
expected in the case of even $d$ only. However, as follows from
eq.~\reef{sczetaSd} and eq.~\reef{oddsczeta} for $s=0$ or
eq.~\reef{oddferzeta} for $s=1/2$, $\zeta_{\BS^d}^{(s)}(0)=0$ for odd
$d$ and so the $\log\delta$ term vanishes as expected. \label{foot7}}
The remaining contributions are precisely those which determine
universal contributions to the \ren entropy
  \be
\log Z_\al^{(s)}={e^{i 2\pi s}\over 2} \  \text{Tr}K_{C_\al}^{(s)}\  \[
\left. { d\zeta_{\BS^d}^{(s)} \over dz}
\right|_{z=0}-\zeta_{\BS^d}^{(s)}(0)\log \delta^2 \] ~.
 \label{spartition3}
 \ee
This expression is readily evaluated using results available in the
literature \cite{camp2,camp}. The $\zeta$-function on the hyperbolic
space $\BH^d$ was computed for the spin-$0$ case in \cite{camp2} and
for spin-${1\over 2}$ case in \cite{camp}. For the scalars, the desired
$\zeta$-functions on $\BS^d$ are then easily obtained from the
hyperbolic ones using a formula given in \cite{camp2} --- see
eq.~\reef{sczetaSd}.

\subsection{\ren entropy for a massive scalar} \label{scalar}

We begin here by evaluating the partition function $Z_\alpha$ for the massive
free scalar field theory described by the action \reef{scalaract}. In this case, the heat kernel
in eq.~\reef{spartition} corresponds to the $D$-dimensional scalar Laplacian on
$\cM_\al$. Separation of the variables leads to eq.~\reef{traces} and further
the heat kernel on $C_\al$ is given by \cite{cone}
 \be
\Tr K_{C_\al}^{(0)}(t)={1\over 12\al}\,(1-\al^2)+\al\ \Tr
K_{\BR^2}^{(0)}(t)
 ~.
 \label{sintKc2}
 \ee
On the other hand, the scalar heat kernel on $\BS^d$ satisfies
 \bea
( -\del_t + \Delta^{(d)})\,K_{\BS^{d}}^{(0)}(t,x,y)&=&0~, \qquad\quad
x,y\in \BS^{d}
 \label{cubes8}\\
K_{\BS^{d}}^{(0)}(0,x,y)&=&\delta(x,y)~,
 \nonumber
 \eea
where $\Delta^{(d)}$ is the scalar Laplacian on $\BS^d$. Of course, due
to the rotational symmetry the heat kernel only depends on the
arc-length between the two points on the sphere. Therefore for
simplicity, we place one of the points at the north pole of the sphere.
With this choice, the heat kernel becomes a function of only the
azimuthal angle $\theta$ and we may replace $\Delta^{(d)}$ by its
radial part.

To solve the resulting equation, we follow the prescription described
in \cite{camp}. That is, we consider the intertwining operator,
$\mathcal{O}=-(2\pi\, \sin\theta)^{-1}\del_\theta$, which relates the
Laplacian on spheres of different dimensions, \ie
 \be
\Delta_\theta^{(d)} \mathcal{O}=\mathcal{O}\(
\Delta_\theta^{(d-2)}+{d-2\over R^2}\)~, \quad
\Delta_\theta^{(d)}=\del_\theta^2+(d-1)\cot\theta\,\del_\theta ~.
 \label{inter9}
 \ee
Here $R$ is the radius of both $\BS^d$ and $\BS^{d-2}$. The overall
constant factor in $\mathcal{O}$ is chosen such that $\mathcal{O}$
relates the delta functions on the two spheres. Hence, for even and odd
dimensions, we have
 \bea
K_{\BS^{2n+1}}^{(0)}(t,\theta)&=&e^{n^2t/R^2}\(-{1\over 2\pi R^2
\sin\theta}\del_\theta\)^n K_{\BS^1}^{(0)}(t,\theta)~,
  \label{sheatSodd}\\
K_{\BS^{2n+2}}^{(0)}(t,\theta)&=&e^{n(n+1)t/R^2}\(-{1\over 2\pi R^2
\sin\theta}\del_\theta\)^{n} K_{\BS^2}^{(0)}(t,\theta)~.
 \label{sheatSeven}
 \eea

For $d=1$, $K_{\BS_1}^{(0)}(t)$ can be readily evaluated using the
method of images. It is given by an infinite sum of  the scalar heat
kernels on $\BR$, which are shifted by integer multiples of $2\pi$ with
respect to each other to maintain periodic boundary conditions for the
scalar field on a circle, namely
 \be
K_{\BS^1}^{(0)}(t,\theta)={1\over\sqrt{4\pi t}}
\sum_{n=-\infty}^{\infty} e^{-{R^2(\theta+2\pi n)^2\over 4t}}
 ~.
 \label{sheatS1}
 \ee
For $d=2$, the heat kernel can be constructed using spherical harmonics
$Y_{lm}(\theta,\phi)$, which correspond to the orthonormal
eigenfunctions of the Laplacian on a unit two-sphere. If one of the
points is taken to the north pole, the result then simplifies to the
following sum
 \be
K_{\BS^2}^{(0)}(t,\theta)={1\over 4\pi
R^2}\sum_{l=0}^{\infty}(2l+1)\,P_{l}(\cos\theta)\, e^{-{l(l+1)\over
R^2} t} ~.
 \label{sheatS2}
 \ee

We will also use the $\zeta$-function formulae given in
eq.~\reef{spartition3}. As shown in \cite{camp2}, $\zeta$-functions on
$\BS^d$ and $\BH^d$ for $d\geq 3$ are related and can be obtained from each other by
means of complex contours. The final result in the case of spin-$0$
case reads
 \bea
 \zeta^{(0)}_{\BS^d}(z)& =& e^{i\pi(z-d/2)} \text{Vol}(\BS^d)
 \Bigg[ {\zeta^{(0)}_{\BH^d}(z) \over \text{Vol}(\BH^d)}
  -i R^{2z-d} e^{-i\pi z} {2^{d-2}\Gamma(d/2)\over \pi^{d/2} } \sin(\pi z)
  \label{sczetaSd}\\
&&\qquad\qquad \times\int_0^\infty{f(ib-y) dy \over
[1+(-1)^de^{2\pi(y-ib)}] [(y-ib)^2+b^2]^z}\Bigg]\quad \text{with\
Re}(z)<1~,
 \nonumber
 \eea
where $\zeta^{(0)}_{\BH^d}$ denotes the scalar $\zeta$-function on
$\BH^d$, which is given by eqs.~\reef{oddsczeta} and \reef{sczeta}.
Following the notation of \cite{camp2}, we also have the following
definitions:
 \bea
b^2&\equiv& -R^2 m_0^2+\frac{(d-1)^2}4 =-R^2 m^2-\xi\,
d(d-1)+\frac{(d-1)^2}4\,,
 \label{deff}\\
f(y)&\equiv&y~{y^2+(d-3)^2/4 \over 4^{d-2}\,\Gamma(d/2)^2}
\prod_{k=-(d-5)/2}^{(d-5)/2}(y+ik)~,
 \nonumber
 \eea
where we have used eq.~\reef{mass0} in the second expression for
$b$.\footnote{This notation is not ideal in the following where we
focus on the limit $mR\gg1$. Hence let us resolve the possible ambiguity
by adding that $b\simeq i\,mR$ in this limit.}  Further for $d=3$ and $4$, the product appearing as the last factor in $f(y)$ should be
omitted.

Even though eq.~\reef{sczetaSd} for $\zeta$-function is only valid for
$\text{Re}(z)<1$, this will be sufficient to compute the \ren entropy in the
present context since according to eqs.~\reef{renyi} and \reef{spartition3}, we
only need to know $\zeta$-function in the vicinity of $z=0$. Note, however,
that general expression which is valid for all values of $z$ can be found in
\cite{camp2}.

The $\zeta$-function approach is similar to dimensional regularization
in that no power law divergences will appear with this method. Hence
the leading contributions to the entanglement entropy of the form given
in eq.~\reef{diverg0} are somewhat obscure in this
framework.\footnote{Of course, this also illustrates the sensitivity of
these contributions to the details of the UV regulator.} In contrast,
evaluating the entanglement (and \ren) entropy using
eq.~\reef{spartition}, where $\delta$ directly cuts off the UV end of
the $t$ integral, produces explicit power law divergences as appear in
eq.~\reef{diverg0}. We illustrate these differences by applying both
approaches in the examples below. Since the form of the heat kernels
and $\zeta$-functions is different in even and odd dimensions, we
consider these cases separately.

\subsubsection*{Odd dimensions}

We start to implement eq.~\reef{spartition} in the special cases $d=1$
and 3 to illustrate how the divergent `area law' and subleading terms
emerge, as well as the universal `area' terms \reef{franks}. We then
consider the $\zeta$-function method \reef{spartition3} to evaluate the
finite contributions to the \ren entropy for general odd $d$.
\newline
\\
{\it $d=1$ ($D=3$):}
\newline
\\
In this case one has to substitute eqs.~\reef{sintKc2} and
\reef{sheatS1} into eqs.~\reef {renyi} and \reef{spartition}
 \bea
S_{\al}^{(0)}&=&{1+\al\over 24\al} \pi^{1/2}R\sum_{n=-\infty}^\infty
\int_{\delta^2}^\infty {dt\over t^{3/2}} \,e^{-t m^2-{(\pi R n)^2\over
t}} \non
&=&{1+\al\over 12\al} \Big( { \text{Vol}(\BS^1) \over
\sqrt{4\pi}\delta} - \log\[ 2\sinh(\pi m R) \] \Big)~.
 \label{ren1}
 \eea
Note that no $\xi$ dependence appears in the expressions above because
the curvature scalar vanishes on $\BS^1$, \ie $m_0=m$ for $d=1$ in
eq.~\reef{mass0}. The divergent term is, of course, the expected `area
law' contribution. It originates from the $n=0$ summand in
eq,~\reef{sheatS1} and thus is independent of the cross-section of the
waveguide geometry, \eg the same term arises in eq.~\reef{sarealaw} for
a hyperbolic waveguide.

In examining the finite contribution above, we first note that it does
not have the simple form expected in eq.~\reef{franks}. However, we
note that the full calculations in \cite{wilczek} resulted in a
similarly complicated expression and the simple universal term only
emerged in the large mass limit. Hence we consider the finite term
above in the limit $mR\gg1$,
 \be
S_{\al,finite}^{(0)}=-{1+\al\over 24\al} \Big(  2\pi R\, m -2 e^{-2\pi
mR} -e^{-4\pi mR}+\cdots \Big)\,.
 \label{simp1}
 \ee
Here we see that the leading term has precisely the form expected in
eq.~\reef{franks} with $\cA_\Sigma=2\pi R$ and $D=3$. We emphasize that
above expression describes the \ren entropy and can be evaluated for
any $\alpha$. The entanglement entropy is recovered by substituting
$\alpha=1$, in which case the pre-factor becomes $-1/12$ precisely
matching the coefficient given in eq.~\reef{coeff} for $D=3$. Here we
also find higher order terms suppressed by exponentials
$\exp(-2\pi\,n\,mR)$. Similar exponential terms were found in
\cite{wilczek} but the precise numerical prefactors do not agree for
the waveguide geometry studied there and for the present cylindrical
waveguide. Of course, these terms only become important when the
Compton wavelength of the scalar is comparable to the size of the
cross-section of the waveguide. Hence it seems these contributions are
probing the topology of the background geometry. In fact, in
eq.~\reef{srenyiodd}, we find that no such exponentials arise when the
cross-section is $\BH^1\simeq \BR^1$.
\newline
\\
{\it $d=3$ ($D=5$):}
\newline
\\
Taking the limit $\theta\rightarrow 0$ in eq.~\reef{sheatSodd}, yields
 \be
K_{\BS^{3}}^{(0)}(t,0)=  {e^{t/R^2} \over  (4\pi t)^{3/2} }
\sum_{n=-\infty}^{\infty} e^{-{\pi^2 R^2 n^2\over t}}\( 1-2\, {\pi^2
R^2 n^2\over t}  \)~.
 \label{heat3}
 \ee
Combining this expression with  eq.~\reef{sintKc2} to form
$K_{\mathcal{M_\al}}^{(0)}$ and in turn, substituting the result into
eqs.~\reef {renyi} and \reef{spartition} yields
 \bea
S_{\al}^{(0)}&=&
{1+\al\over 24\al}\,\text{Vol}(\BS^3)\ \int_{\delta^2}^{\infty}
{dt\over t} e^{-t\big(m^2+6\xi/R^2\big)}K_{\BS^{3}}^{(0)}(t,0)
  \label{ren3}\\
&=& {1+\al\over 12\,\al} \, {\text{Vol}(\BS^3) \over (4\pi)^{3/2}}\Big[
{1\over 3\,\delta^{3} } -\Big(m^2+{6\xi-1\over R^2}\Big){1\over \delta}
\Big] +  {1+\al\over 96\,\pi^2\,\al} \, g^{(0)}\Big(2\pi\sqrt{(m
R)^2+6\xi-1}\Big)~,
 \nonumber
 \eea
with $\text{Vol}(\BS^3)=2\pi^2 R^3$ and
 \be
g^{(0)}(x)={1\over 6}
x^3+x^2\log(1-e^{-x})-2\,\text{Li}_3(e^{-x})-2\,x\,\text{Li}_2(e^{-x})\,,
 \label{ren3a}
 \ee
where $\text{Li}_3(z)$ and $\text{Li}_2(z)$ are the standard
polylogarithms. In eq.~\reef{ren3}, we see the expected area law
contribution proportional to $\text{Vol}(\BS^3)/\delta^3$, as well as a
subleading divergences proportional to $1/\delta$. Focusing on the
finite contribution in the limit $mR\gg1$, we find
 \bea
S_{\al,finite}^{(0)}&=&{1+\al\over 144\pi\,\al}\ \cA_\Sigma\,
\left(m^3+\frac{3(6\xi-1)}2\,\frac{m}{R^2}+\frac{3(6\xi-1)^2}8\,\frac{1}{m\,R^4}+\cdots
 \right.\non
&&\left.\qquad\qquad\qquad\qquad-\frac{3}\pi\,\frac{m^2}R\, e^{-2\pi
mR} +\cdots\ \  \right)
 \label{simp3}
 \eea
where $\cA_\Sigma=\text{Vol}(\BS^3)=2\pi^2 R^3$ is the area of the
entangling surface. The leading term in this expansion has the form
expected from eq.~\reef{franks} and setting $\al=1$ to recover the
entanglement entropy, the prefactor becomes $1/(72\pi)$ which precisely
matches the coefficient given in eq.~\reef{coeff} with $D=5$. There are
two classes of subleading terms: First there is an expansion in powers
of $1/(mR)^2$, which produces terms where the prefactor becomes
$\cA_\Sigma m^3/(mR)^{2n}$. These contributions have precisely the form
expected for the curvature terms described in eq.~\reef{univEE}.
Second, there are contributions with exponential factors
$\exp(-2\pi\,n\,mR)$, similar to those found with $d=1$. As explained
above, it appears that these contributions probe the topology of the
waveguide geometry. Note that, as illustrated in eq.~\reef{simp3}, some
of these exponential contributions contain odd\footnote{Here, we mean
odd powers of $1/R$ multiplying $\cA_\Sigma$, which itself contains a
factor of $R^3$.} powers of $1/R$, which emphasizes that these terms
cannot be given a simple geometric interpretation, as in
eq.~\reef{univEE}.
\newline
\\
{\it General odd $d\geq 3$:}
\newline
\\
As already noted in footnote \ref{foot7}, the logarithmic divergence in
eq.~\reef{spartition3} vanishes for odd $d$ since
$\zeta^{(0)}_{\BS^d}(0)=0$. Hence, evaluating the finite term in the
\ren entropy using eqs.~\reef{renyi}, \reef{spartition3},
\reef{sintKc2}  and \reef{sczetaSd} yields:
 \be
 S_{\al,finite}^{(0)}=(-1)^{d+2\over 2}{1+\al\over 24\al}\,\text{Vol}(\BS^d)
  \({1\over \text{Vol}(\BH^d) }\left. {d\zeta_{\BH^d}^{(0)} \over dz} \right|_{z=0}
  -i {2^{d-2}\Gamma(d/2)\over \pi^{d-2\over 2} R^{d} }
 \int_0^\infty{f(ib-y) dy \over 1-e^{2\pi(y-ib)}}\)~.
 \label{generalodd1}
 \ee
Substituting eq.~\reef{oddsczeta} for $\zeta_{\BH^d}^{(0)}$ above, then
produces
 \bea
S_{\al,finite}^{(0)}&=&(-1)^{d\over 2}{1+\al\over
24\al}\,{\text{Vol}(\BS^d)\over \pi^{d-2\over 2} R^{d}}
  \Bigg(
   { 1\over 2^{d-1} \Gamma(d/2)}
  \sum_{k=0}^{(d-1)/2} g_{k,d}^{(0)}\,
   b^{2k+1} {\sec(k\pi)\over 2k+1}
 \label{srenyioddS}\\
 &&\qquad\qquad\qquad\qquad\qquad\qquad
  +i \, 2^{d-2}\Gamma(d/2)
 \int_0^\infty{f(ib-y) dy \over 1-e^{2\pi(y-ib)}}\Bigg)~,
 \nonumber
 \eea
 where $g_{0,3}^{(0)}=0$, $g_{1,3}^{(0)}=1$ and $g_{k,d}^{(0)}$ for odd $d\geq 5$ are defined by
 \be
\Big[x^2+\Big({d-3\over 2}\Big)^2\Big]\prod_{j=0}^{(d-5)/2}
(x^2+j^2)=\sum_{k=0}^{(d-1)/2}g_{k,d}^{(0)}\,x^{2k} ~.
 \label{defg}
 \ee
From this expression, it is useful to note that
 \be
g_{\frac{d-1}2,d}^{(0)}=1\quad{\rm and}\quad g_{\frac{d-3}2,d}^{(0)}=
\frac1{24}(d-1)(d-2)(d-3) \,,
 \label{useg}
 \ee
as well as $g_{0,d}^{(0)}=0$. We may then use these expressions to
expand $S_{\al,finite}^{(0)}$ in the limit $mR>>1$ and in doing so, we
find
 \be
S_{\al,finite}^{(0)}={1+\al\over 24\al}\,{(-1)^{D-1\over 2}\pi\over
(4\pi)^{D-2\over 2} \Gamma\big({D/2}\big)}
\cA_\Sigma\Big(m^{D-2}+{(D-2)^2(D-3)(6\xi-1)\over 12}{m^{D-4}\over
R^2}+\ldots\Big) \label{renfinSodd}
 \ee
where as before $\cA_\Sigma=\text{Vol}(\BS^d)$ is the area of the entangling
surface. Of course, setting $D=5$, eq.~\reef{renfinSodd} simply produces the
first two terms in eq.~\reef{simp3}. For general $D$, we may set $\al=1$ to
recover the entanglement entropy and we see that the leading term above is the
precisely the area term expected in eqs.~\reef{franks} and \reef{coeff}.  The
next to leading term in eq.~\reef{renfinSodd} introduces a new universal
contribution to the entanglement entropy which matches the form shown in
eq.~\reef{univEE} with $n=1$. This contribution can be interpreted as
 \be
S_\mt{univ}=\frac{D-2}{2}\,\left(\xi -\frac16 \right)\gamma_{D,scalar}\,
\int_\Sigma \!d^{D-2}\!\sigma\,\sqrt{h}\ \cR(h)\ m^{D-4}
 \label{firstR}
 \ee
where $\cR(h)$ is the Ricci scalar of the metric induced on the entangling
surface and the coefficient $\gamma_{D,scalar}$ is precisely that given in
eq.~\reef{coeff}. Note that we should only consider this term for odd $D\ge5$.
Of course, there are several other independent curvature terms which could in
general contribute at this order but with an appropriate choice of basis, the
extra terms all vanish for the waveguide geometry studied here --- see section
\ref{discuss} for further discussion.

\subsubsection*{Even dimensions}

Following our discussion of odd $d$, we first consider the special
value $d=2$ here and evaluate all UV divergences using
eqs.~\reef{renyi} and \reef{spartition}. Then for general even $d\geq
4$, we apply the approach of $\zeta$-function regularization. This
method eliminates power law divergences, while keeping the universal
terms, \ie logarithmic divergences, as well as finite contributions to
the \ren entropy.
\newline
\\
{\it $d=2$ ($D=4$):}
\\
\newline
In the limit of coincident points, we get from eq.~\reef{sheatS2}
 \bea
 K_{\BS^2}^{(0)}(t)={1\over 4\pi R^2}\sum_{l=0}^{\infty}
(2l+1) e^{-{l(l+1)\over R^2} t}\, .
 \eea
Applying the Euler-Maclaurin formula, \ie
 \be
  \sum_{l=0}^{\infty}F(l)\simeq\int_0^{\infty}F(x)+{F(0)+F(\infty)\over 2}+
  \sum_{l=1}^{\infty}{B_{2l}\over (2l)!}\( F^{(2l-1)}(\infty)-F^{(2l-1)}(0) \)~,
 \ee
leads to the following expansion
 \be
 K_{\BS^2}^{(0)}(t)={1\over 4\pi t}+{1\over 12\pi R^2}+\mathcal{O}(t).
 \ee
Combining this expression with  eq.~\reef{sintKc2} yields
$K_{\mathcal{M_\al}}^{(0)}$ as in eq.~\reef{sepheat}. Then substituting
the result into eqs.~\reef {renyi} and \reef{spartition} yields
 \bea
S_{\al}^{(0)}&=&{1+\al\over 24\al}\,\text{Vol}(\BS^2)\,
\int_{\delta^2}^{\infty} {dt\over t}
e^{-t\big(m^2+2\xi/R^2\big)}K_{\BS^2}^{(0)}(t)
  \non
&=&{1+\al\over 48\pi\al}\,\cA_\Sigma \,\[ {1 \over 2 \, \delta^{2} }
+\Big(m^2+ {6\xi -1 \over 3 R^2} \Big) \log(m\delta)+...\]
  \label{sren4dS}
  \eea
where $\cA_\Sigma=\text{Vol}(\BS^2)=4\pi R^2$ and ellipsis denotes
finite terms. Upon setting $\al=1$, we recover the entanglement entropy
and the first term is recognized as the standard `area law'
contribution. The logarithmic contribution proportional to $m^2$
matches the area term given in eq.~\reef{franks} with $D=4$. With
$\xi=1/6$ and $m^2=0$, the theory \reef{scalaract} under consideration
becomes a conformal scalar. In this case, one can verify that the
logarithmic contribution above (which vanishes) matches the expected
result from eq.~\reef{WaldS9} for a conformal scalar --- see section
\ref{discuss} for further discussion.
\newline
\\
{\it General even $d\geq 4$:}
\newline
\\
According to eqs.~\reef{renyi} and \reef{spartition3} for even $d$, the
universal term  is proportional to $\zeta^{(0)}_{\BS^d}(0)$. We can
evaluate the latter using eq.~\reef{sczetaSd}. Further, we see that
since $\sin(\pi z)$ vanishes at the origin, the latter expression
simplifies to
 \be
\zeta^{(0)}_{\BS^d}(0)=(-1)^{d/2}  {\text{Vol}(\BS^d) \over
\text{Vol}(\BH^d)}\zeta^{(0)}_{\BH^d}(0)~.
 \ee
Hence, using eq.~\reef{sczeta} for the scalar $\zeta$-function on
$\BH^d$, we find that universal contribution is given by
 \be
S_{\al, univ}^{(0)}={1+\al \over 12\al} { (-1)^{d+2\over
2}\text{Vol}(\BS^{d})\over (4\pi)^{d/2} \Gamma(d/2)R^{d}}
\sum_{k=0}^{(d-2)/2} h_{k,d}^{(0)}\[   {(-b^2)^{k+1} \over
k+1}-4\int_0^\infty {x^{2k+1}\over e^{2\pi x}+1}dx \] \log(m \delta) \,
, \label{srenyievenS}
 \ee
where $h_{0,4}^{(0)}=1/4$, $h_{1,4}^{(0)}=1$ and $h_{k,d}^{(0)}$ for
even $d\geq 6$ are defined by
 \be
\Big[x^2+\Big({d-3\over 2}\Big)^2\Big]\prod_{j=1/2}^{(d-5)/2}
(x^2+j^2)=\sum_{k=0}^{(d-2)/2}h_{k,d}^{(0)}x^{2k}
 ~.\label{defh}
 \ee
Given this definition, it is useful to note that
 \be
h_{\frac{d-1}2,d}^{(0)}=1\quad{\rm and}\quad h_{\frac{d-3}2,d}^{(0)}=
\frac1{24}(d-1)(d-2)(d-3) \,.
 \label{useh}
 \ee
Using the latter two expressions in an expansion of
$S_{\al,finite}^{(0)}$ in the limit $mR>>1$ yields
 \be
S_{\al, univ}^{(0)}={1+\al \over 12\al} { (-1)^{D\over
2}\cA_\Sigma\over (4\pi)^{D-2\over 2} \Gamma(D/2)}
\Big(m^{D-2}+{(D-2)^2(D-3)(6\xi-1)\over 12}{m^{D-4}\over
R^2}+\ldots\Big) \log(m \delta) \, .
 \label{exprenSev}
 \ee
As usual $\cA_\Sigma=\text{Vol}(\BS^d)$ is the area of the entangling surface.
Note that setting $D=4$ (\ie d=2), eq.~\reef{exprenSev} reproduces the
universal term calculated above in eq.~\reef{sren4dS}. Again with $\al=1$, the
above reduces to the entanglement entropy and we see that the leading term is
the precisely the area term expected in eqs.~\reef{franks} and \reef{coeff}.
Similar to the discussion for odd $d$, The next to leading term in
eq.~\reef{exprenSev} introduces a new universal contribution to the
entanglement entropy which again matches the form shown in eq.~\reef{univEE}
with $n=1$. We can write this contribution as
 \be
S_\mt{univ}=\frac{D-2}{2}\,\left(\xi-\frac16
\right)\gamma_{D,scalar}\,\int_\Sigma \!d^{D-2}\!\sigma\, \sqrt{h}\ \cR(h)\
m^{D-4}\,  \log(m \delta)
 \label{firstRx}
 \ee
where $\cR(h)$ is the Ricci scalar of the metric induced on the entangling
surface and the coefficient $\gamma_{D,scalar}$ is given in eq.~\reef{coeff}.
Here we should only consider this term for even $D\ge4$. Of course, this
expression is reminiscent of eq.~\reef{firstR} for the case of odd $d$.

If we evaluate the entire expression \reef{srenyievenS} for $d=4$ ($D=6$), we
obtain
 \be
S_{\al, univ}^{(0)}(d=4)=-{1+\al \over \al} { \text{Vol}(\BS^{4})\over
192\,\pi^2} \[ {1\over 2}m^4+ 2(6\xi-1) {m^2\over R^2}
+\(72\xi^2-24\xi+{29\over 15}\){1\over R^4}
  \] \log(m\delta)
 ~.
 \label{6Dscal}
 \ee
This example illustrates that the curvature contributions in
eq.~\reef{univEE} extend up to $n=d/2$ for even $d$, as can also be
seen by directly examining eq.~\reef{srenyievenS}. In contrast to the
case of odd $d$, these two equations also show that there are no
exponentially suppressed terms in the universal contribution for even
$d$.

\subsection{\ren entropy for a massive fermion} \label{fermion}

In this section, we construct the partition function for a spin-${1\over 2}$
field living on the Euclidean manifold $\cM_\al=C_\al\times \BS^{d}$ and use
this result to evaluate the corresponding \ren entropy. Our spinor notation is
reviewed in Appendix \ref{appx:sepvar}, whereas the action under consideration
is given by
 \be
S(\psi, \bar\psi)= \int_{\cM_\al}
\big(\bar\psi\slashed\nabla\psi+m\,\bar\psi\psi\big)~. \label{fermact}
 \ee
In this case, the massless wave operator appearing in the heat kernel
\reef{spartition} is
 \be
\slashed\nabla\cdot\slashed\nabla^{\dag}=- \slashed\nabla^2\,,
 \label{iterate}
 \ee
which we refer to as the `iterated' Dirac operator, following \cite{camp}.
Since the Dirac operator is a nondiagonal matrix, the separation of variables
in eq.~\reef{sepheat} is not obvious and so we prove that this equation still
holds here in Appendix \ref{appx:sepvar}. The argument there rests on the
structure of the heat kernel for the iterated Dirac operator on $\BS^d$, which
we review next.

Let us first consider the case of odd $d$ and further for simplicity, let us
assume that one of the points coincides with the north pole of $\BS^d$. In this
case, the heat kernel reduces to \cite{camp}
 \be
K_{\BS^{2j+1}}^{(1/2)}(t,y)=\hat U(y) \cos\frac{\theta}{2}  \left({1\over 2\pi
R^2}\frac{\del} {\del \cos \theta}\right)^j
\Big(\cos\frac{\theta}{2}\,\Big)^{-1} \sum_{n=-\infty}^{+\infty}(-1)^n {
e^{-\frac{\theta_n^2 R^2}{4t}} \over (4\pi t)^{1/2}}\,,
 \label{fermheatSodd}
 \ee
where $y$ is an arbitrary point on the sphere. The angle of latitude for this
point is designated as $\theta$ and then $\theta_n=\theta+2\pi n$. Finally
$\hat U(y)$ is the spinor matrix which parallel propagates a spinor from the
given point $y$ to the north pole. Similarly, in the case of even $d$, the heat
kernel becomes \cite{camp}
 \be
K_{\BS^{2j+2}}^{(1/2)}(t,y)=\hat U(y) \cos\frac{\theta}{2}  \left({1\over 2\pi
R^2}\frac{\del} {\del \cos \theta}\right)^j
\Big(\cos\frac{\theta}{2}\,\Big)^{-1} f_{\BS^2}^{(1/2)}(\theta, t),
 \label{fermheatSeven}
 \ee
where
 \be
f_{\BS^2}^{(1/2)}(\theta, t)={\sqrt{2}\,R \over (4\pi t)^{3/2}\cos(\theta/2) }
\sum_{n=-\infty}^{+\infty} \int_\theta^\pi {\phi_n \,\cos {\phi \over 2} \over
\sqrt{\cos\theta-\cos\phi}} \, e^{-{R^2\phi_n^2\over 4t} } d\phi
  ~,
 \ee
with $\phi_n=\phi+2\pi n$.

The structure of the spinor matrix $\hat U(y)$ can be found in
\cite{camp}\footnote{We also refer the interested reader to
eq.~\reef{propell}.} but these details will be not important here because we
are only interested in the limit of coincident points. In this limit, $\hat
U(y)$ simply reduces to an identity matrix. We might note that if we were
considering the heat kernel on $\BS^d$ alone, the dimension of this identity
matrix would be $2^{\lfloor d/2\rfloor}$, \ie the dimension of Dirac spinors in
$d$ dimensions. Of course, here $\BS^d$ is part of the larger manifold
$\cM_\al$ and so the dimension of $\hat U(y)$ is actually $2^{\lfloor
D/2\rfloor}$. However, following the conventions introduced in Appendix
\ref{appx:sepvar}, we treat the spinor trace on the right-hand side of
eq.~\reef{traces} as though we separately tracing over the spinor spaces of the
two component manifolds. That is, we calculate the two spinor heat kernels on
$C_\alpha$ and $\BS^d$ separately and then simply take their product in
eq.~\reef{traces}.

The spinor heat kernel on the cone $C_\al$ is readily evaluated as
\cite{kabat,jackiw}
 \be
\Tr K_{C_\al}^{(1/2)}(t) = -{1 \over 12\al} \Big(   1- \alpha^2 \Big)+\al\,\Tr
K_{\BR^2}^{(1/2)}(t)\,.
 \label{intKc2fer2}
 \ee
As noted above, this result accounts for the trace over the two-dimensional
spinor indices on the cone $C_\alpha$.

Unfortunately, in the present case, we are unable to apply the
$\zeta$-function approach, which would have allowed a systematic
evaluation of the universal contributions to the \ren entropy for
general $d$. In particular, while the spinor $\zeta$-function is known
for the hyperbolic space $\BH^d$ \cite{camp}, the spin-$\frac12$
counterpart of eq.~\reef{sczetaSd} relating these to the desired
$\zeta$-functions on $\BS^d$ is unavailable. Hence, in the following,
we limit ourselves to considering a few special cases, \ie $d=1$, 2 and
3. In each case, the \ren entropy is determined by simply substituting
eq.~\reef{fermheatSodd} or \reef{fermheatSeven}, along with
eq.~\reef{intKc2fer2} into eqs.~\reef {renyi} and \reef{spartition}.
Given these expressions above, the generalization of the following
results to higher dimensions would be straightforward.

Let us add that we consider fermions on the hyperbolic waveguide
$C_\al\times\BH^d$ in Appendix \ref{renyiSd}. In this case, we can follow the
$\zeta$-function approach using the results of \cite{camp}. This allowed us to
produce results for the \ren entropy of spin-$\frac12$ fields for general $d$
and in particular, gave the general coefficients appearing in
eqs.~\reef{coeff2} and \reef{coeff8x}.
\newline
\\
{\it $d=1$ ($D=3$):}
\\
\newline
In this case using eq.~\reef{fermheatSodd}, we find
 \bea
S_{\al}^{(1/2)}&=&{1+\al\over 24\al} \pi^{1/2}R\sum_{n=-\infty}^\infty
(-1)^n\int_{\delta^2}^\infty {dt\over t^{3/2}}
 \,e^{-t m^2-{(\pi R n)^2\over t}}
 \nonumber\\
&=&{1+\al\over 12\al} \Big( { \text{Vol}(\BS^1) \over
\sqrt{4\pi}\delta} - \log\[ 2\cosh(\pi m R) \] \Big)~.
 \label{ferenS1}
 \eea
In the large mass limit (\ie $mR\gg1$), the finite contribution becomes
 \be
S_{\al,finite}^{(1/2)}=-{1+\al\over 24\al}\big(2\pi R\, m +2e^{-2\pi m
R}-e^{-4\pi m R}+\cdots\big)~.
 \label{ferenS1univ}
 \ee
Hence the leading term has precisely the form expected in
eq.~\reef{franks} with $\cA_\Sigma=2\pi R$ and $D=3$. The entanglement
entropy is recovered by substituting $\alpha=1$, in which case the
pre-factor matches that given in eq.~\reef{coeff2} for $D=3$. Note that
this area term is accompanied by higher order exponential terms,
similar to those found in eq.~\reef{simp1}.
\newline
\\
{\it $d=2$ ($D=4$):}
\\
\newline
In the limit of coincident points, eq.~\reef{fermheatSeven} yields
 \be
 K_{\BS^2}^{(1/2)}(t)= f_{\BS^2}^{(1/2)}(0, t)={R \over 2(\pi t)^{3/2}   }
  \sum_{n=-\infty}^{+\infty}\int_0^{\pi\over 2}
  (\phi+\pi n) \,\cot\!\phi\ e^{-{R^2(\phi+\pi n)^2\over t} } d\phi
  ~. \label{coincide2}
 \ee
We may note that the integrand is regular at the lower bound. While $\cot\phi$
has a simple pole at $\phi=0$, the coefficient of this pole is proportional to
$\sum_{n=-\infty}^{+\infty} n\,e^{-{(R\pi n)^2\over t} }$ which vanishes
because the summand is odd in $n$. Redefining the integration variable
$\phi\rightarrow \phi/\sqrt{t}$, we have
 \bea
 K_{\BS^2}^{(1/2)}(t)={R \over 2\,\pi^{3/2} t^{1/2}  }
  \sum_{n=-\infty}^{+\infty}\int_0^{\pi\over 2\sqrt{t}}
  \Big(\phi+{\pi n\over \sqrt{t}} \Big) \,\cot(\sqrt{t}\phi)
  \, e^{-R^2\big(\phi+{\pi n\over \sqrt{t}}\big)^2 } d\phi
  ~.
 \eea
From this expression, it is obvious that the UV divergences in
eq.~\reef{spartition} come entirely from the $n=0$ term. Any terms with $n\ne0$
contain an exponential factor $e^{-\pi^2R^2n^2/t}$ which smoothes out any
potential singularities at $t=\delta^2$. Therefore, to extract the structure of
the UV divergences, and in particular, the $\log(m\delta)$ contribution, it is
enough to examine only the $n=0$ term, \ie
 \bea
K_{\BS^2}^{(1/2)}(t)={R \over 2\,\pi^{3/2} t^{1/2}  } \int_0^{\pi\over
2\sqrt{t}}\phi \,\cot(\sqrt{t}\phi) \, e^{-R^2\phi^2 } d\phi\,.
 \label{horsex}
 \eea
The remaining terms only contribute to the finite part of the \ren
entropy.

Now we expand the integrand in eq.~\reef{horsex} in the vicinity of $t=0$ and
find that it is sufficient to keep only first two terms since the rest do not
lead to singularities in the limit $\delta\rightarrow 0$
 \bea
K_{\BS^2}^{(1/2)}(t)={1 \over 2\,\pi^{3/2} t  } \int_0^{\infty}
(1-{t\over 3R^2} \, x^2+\ldots) \, e^{-x^2 } dx +\ldots={1\over 4\pi t}
- {1\over 24\pi R^2}+\ldots
 \eea
Combining this result with  eq.~\reef{intKc2fer2} to form
$K_{\mathcal{M_\al}}^{(1/2)}$ and substituting the result into eqs.~\reef
{renyi} and \reef{spartition}, we find
 \bea
S_{\al}^{(1/2)}&=&{1+\al\over 12\al} \,\text{Vol}(\BS^2) \,
\int_{\delta^2}^{\infty} {dt\over t} e^{-t m^2}K_{\BS^2}^{(1/2)}(t)
  \nonumber\\
&=&{1+\al\over 48\pi\al}\,\cA_\Sigma \,\( {1 \over \, \delta^{2} }
+\Big(2 m^2 + {1 \over 3 R^2} \Big) \log(m\delta)+...\)
  \label{ferren4dS}
  \eea
where $\cA_\Sigma=\text{Vol}(\BS^2)=4\pi R^2$ and ellipsis denotes finite
terms. For $\al=1$ the first term represents the standard `area law' in the
entanglement entropy, whereas the next term proportional to $m^2$ is precisely
the $D=4$ case of eqs.~\reef{franks} and \reef{coeff2}. Similarly, the term
proportional to $1/R^2$ matches eqs.~\reef{newX} and \reef{coeff8x} with $D=4$.
\newline
\\
{\it $d=3$ ($D=5$):}
\\
\newline
Taking the limit $\theta\rightarrow 0$ with $d=3$ in
eq.~\reef{fermheatSodd}, yields
 \be
 K_{\BS^{3}}^{(1/2)}(t,0)=  {1 \over  (4\pi t)^{3/2} } \sum_{n=-\infty}^{\infty} (-1)^n
 e^{-{(\pi R n)^2\over t}}\( 1 -{t \over 2 R^2} \)~.
 \ee
Combining the above with eq.~\reef{intKc2fer2} to form
$K_{\mathcal{M_\al}}^{(1/2)}$ and substituting the result into
eqs.~\reef {renyi} and \reef{spartition}, yields
 \bea
S_{\al}^{(1/2)}&=&{1+\al\over 12\al} \,
\text{Vol}(\BS^3)\int_{\delta^2}^{\infty} {dt\over t} e^{-t
m^2}K_{\BS^{3}}^{(1/2)}(t,0)
  \non
&=& {1+\al\over 6\,\al} \, {\text{Vol}(\BS^3) \over (4\pi)^{3/2}}\Big[
{1\over 3\,\delta^{3} }
  -\Big(m^2+{1\over 2 R^2}\Big){1\over \delta}  \Big]
  +  {1+\al\over 48\,\pi^2\,\al} \, g^{(1/2)}(2\pi R m)~,
  \label{ferenS3}
 \eea
where
 \be
g^{(1/2)}(x)={x^3\over 6} +{\pi^2\over
2}x+\pi^2\log(1+e^{-x})+\,\text{Li}_3(-e^{-x})+\,x\,\text{Li}_2(-e^{-x})~.
 \label{long}
 \ee
Expanding the finite contribution above in the limit $mR\gg1$, we find
 \be
S_{\al,finite}^{(0)}={1+\al\over 72\pi\,\al}\ \cA_\Sigma\,
\left(m^3+\frac{3}4\,\frac{m}{R^2}
-\frac{3}{2\pi^2}\,\frac{m}{R^2}\, e^{-2\pi
mR} +\cdots\ \  \right)
\label{ferenS3univ}
 \ee
where
$\cA_\Sigma=\text{Vol}(\BS^3)$. Setting $\al=1$ to recover the entanglement
entropy, the pre-factor becomes $1/(36\pi)$. In this case, the leading term in
this expansion is the area term \reef{franks} with precisely the coefficient
given in eq.~\reef{coeff2} for $D=5$. Also, the next term matches
eqs.~\reef{newX} with the coefficient given by eq.~\reef{coeff8x} for $D=5$.
Further, the above expansion also reveals contributions with exponential
factors $\exp(-2\pi\,n\,mR)$, similar to those found previously for odd
dimensions.

\section{A calculation at strong coupling} \label{strong}

In this section, we are going to use gauge/gravity duality to study the
universal `area' contribution to the entanglement entropy of $\caln=2^*$ gauge
theory \cite{pw,bpp} at strong coupling. The latter is a massive deformation of
the four-dimensional $\caln=4$ super-Yang-Mills (SYM) theory, which is commonly
studied in the AdS/CFT correspondence. So let us begin by giving the field
theoretic description of the mass terms which appear in this context. The
$\caln=4$ SYM theory includes a gauge field $A_\mu$, four Majorana fermions
$\psi_a$ and three complex scalars $\phi_i$, all of which are in the adjoint
representation of the $U(N)$ gauge group. Now there are two independent `mass'
terms which can be used to deform the SYM theory \cite{opera}
 \beq
\delta \call= -2\,\int d^4x\,\left[ \,m_b^2\,\calo_2
+m_f\,\calo_3\,\right]
 \label{massterm}
 \eeq
where
 \beqa
\calo_2&=&\frac13 {\tr}\left(\, |\phi_1|^2 + |\phi_2|^2 - 2\,|\phi_3|^2
\,\right)\,,
 \label{massb}\\
\calo_3&=& -{\tr}\left( i\,\psi_1\psi_2 -\sqrt{2}g_\mt{YM}\,\phi_3
[\phi_1,\phi_1^\dagger] +\sqrt{2}g_\mt{YM}\,\phi_3
[\phi_2^\dagger,\phi_2] + {\rm h.c.}\right)
 \label{massf}\\
&&\qquad\qquad +\frac23 m_f\, {\tr}\left(\, |\phi_1|^2 + |\phi_2|^2 +
|\phi_3|^2\, \right)\,.
 \nonumber
 \eeqa
The $\caln=2^*$ gauge theory results when we set $m_b=m_f$ and the effect is to
give mass to the $\caln=2$ hypermultiplet comprised of $\phi_{1,2}$ and
$\psi_{1,2}$. At this point, we note that the dimension-two operator $\calo_2$
contains an unstable mass term for the scalar $\phi_3$ --- a typical
characteristic of such superconformal primary operators. This negative
mass-squared contribution from $\calo_2$ is precisely canceled by the positive
contribution in $\calo_3$ when $m_b=m_f$ and hence $\phi_3$ is left massless in
the supersymmetric theory. In the following, we will not restrict our attention
to the supersymmetric theory and instead we set $m_b$ and $m_f$ to independent
values. In this case, one may then worry that the resulting theory is unstable,
\ie when $m_b>m_f$. Determining the end-point of this instability is the
question of understanding the infrared behaviour of the RG flow induced by the
mass terms.\footnote{The IR theory can be stabilized by introducing a finite
temperature, \eg \cite{n22,n23}. However, such a modification of the IR state
would again not modify the coefficient of the logarithmic contribution in the
entanglement entropy \cite{relevant}.} However, recall that the universal
contribution to the entanglement entropy appears with a logarithmic dependence
on the UV cut-off, $\log\delta$, as shown in eq.~\reef{franks}, since we are
studying a four-dimensional gauge theory here. Furthermore, as demonstrated in
\cite{relevant} and as we will explicitly see below, the coefficient of this
logarithmic term is only determined by the UV properties of the RG flow and is
completely insensitive to the IR details. Hence, any such instability is of no
consequence to the following calculations.

The dual holographic theory consists of five-dimensional Einstein gravity
coupled to a pair of scalars, $\alpha$ and $\chi$, as described by the
following action \cite{pw,bpp}:
\begin{equation}
\begin{split}
I_5=&\,
\int_{\calm_5} d\xi^5 \sqrt{-g}\ \call_5=\frac{1}{16\pi G_5}\,
\int_{\calm_5} d\xi^5 \sqrt{-g}\left[R-3 (\del\a)^2- (\del\chi)^2-
 \calv(\a,\chi) \right]\,,
\end{split}
\label{action5o}
\end{equation}
where the potential takes the form
\begin{equation}
\calv(\a,\chi)=-\frac{4}{L^2}e^{-2\alpha} -\frac{8}{L^2}e^{\alpha}\cosh\chi
+\frac1{L^2}\,e^{4\alpha}\sinh^2\chi\,.
 \label{pp}
\end{equation}
Given the above expression, it is trivial to show $\calv(\a=0,\chi=0)=-12/L^2$
and so we see that $L$ corresponds to the curvature scale of the AdS$_5$ vacuum
solution. As it will be needed below, we also note that in the present
conventions, Newton's constant is given by
\begin{equation}
G_5\equiv \frac{\pi L^3}{2 N^2}\,,
\label{g5}
\end{equation}
where $N$ is the rank of the $U(N)$ gauge group in the boundary theory. As is
well-known, the AdS/CFT correspondence relates the asymptotic boundary
behaviour of the scalars to the couplings and expectation values of the dual
operators in the boundary theory. Here, $\a$ is dual to the `{\it bosonic}'
mass coupling $m_b^2$ and the corresponding operator $\calo_2$, given in
eq.~\reef{massb}. Similarly, $\chi$ is dual to the `{\it fermionic}' mass $m_f$
and the operator $\calo_3$ in eq.~\reef{massf}.

In this holographic context, we follow the now standard approach to calculating
entanglement entropy \cite{rt2,rt1}. That is, given a spatial region $V$ in the
boundary theory, the entanglement entropy between this region and its
complement is given by the following expression evaluated in the bulk
spacetime:
 \be
S(V) = \frac{1}{4G_5}\ \mathrel{\mathop {\rm
ext}_{\scriptscriptstyle{\partial\m\sim \partial V}} {}\!\!}
\left[A(\m)\right]\,.
 \label{define}
 \ee
Here $\m$ is a (three-dimensional) bulk surface extending out to asymptotic
infinity such that its asymptotic boundary $\partial \m$ matches the
`entangling surface' $\partial V$ in the boundary geometry. The symbol `ext'
indicates that one should extremize the area over all such surfaces
$\m$.\footnote{If eq.~\reef{define} is calculated in a Minkowski signature
background, the extremal area is only a saddle point. However, if one first
Wick rotates to Euclidean signature, the extremal surface will yield the
minimal area.}

Following \cite{opera} (see also \cite{n22,n23}), we adopt the following ansatz
for the bulk solution
\begin{eqnarray}
ds^2&=&
\frac{L^2}{z^2}\left[e^{2 A(z)}(-B(z)^2 dt^2+d \vec{x}^2)+ dz^2\right]
 \label{anzata}\\
\alpha&=&\alpha(z)\,,\qquad \chi=\chi(z)\,.\nonumber
\end{eqnarray}
This solution reduces to simply AdS$_5$ with $A=0=B=\alpha=\chi$. In general,
turning on the bulk scalars induces a holographic RG flow which is encoded in
the metric function $A(z)$. The function $B(z)$ provides a potential blackening
factor in the time component of the metric, which also allows the above ansatz
to describe finite temperature situations.

In order to identify the $\log\delta$ term in the entanglement entropy, we will
only need to consider the asymptotic or small $z$ region of the bulk solution,
which describes the UV behaviour of the dual gauge theory. The desired
asymptotic solution was found in \cite{opera} and takes the form
\begin{eqnarray}
\alpha(z)&=& z^2 ( \alpha_{0,1} \log(z) + \alpha_{0,0})+O(z^4 \log^2 z)\,,
  \nonumber \\
\chi(z)&=&z\,  \chi_{0,0} +z^3 \left ( \frac{4}{3} \chi_{0,0}^3 \log z + \chi_{2,0}
 \right)+ O(z^5 \log^2 z)\,, \label{azymp}\\
 A(z)&=&-\frac{\chi_{0,0}^2}{3}\,z^2+ O(z^4 \log^2 z)\,,\nonumber \\
 B(z)&=&1+B_{4,0}\, z^4 + O(z^6)\,. \nonumber
 \end{eqnarray}
Of particular importance, the coefficients, $\alpha_{0,1}$ and $\chi_{0,0}$,
correspond to the field theory masses \cite{opera}\footnote{With respect to
conventions in \cite{opera}, we have already extracted the $L$ dependence in
the metric.}
 \begin{equation}
 \alpha_{0,1}=\dfrac{m_b^2}{6}\qquad{\rm and}
\qquad \chi_{0,0}=\dfrac{m_f}{2}
\,. \label{hum3}
\end{equation}
Similarly, the coefficients $\alpha_{0,0}$ and $\chi_{2,0}$ are related to the
expectation values of $\calo_2$ and $\calo_3$, respectively, while $B_{4,0}$
yields the energy density. We might add that the exact solution describing the
supersymmetric flow (with $m_f=m_b=m$) is known \cite{pw} and the corresponding
asymptotic expansion then becomes \cite{n22,n23},
\begin{eqnarray}
\alpha(z)&=&  \frac{m^2 z^2}{6} \left (\log (\frac{m z}{2})+\frac12 \right )+
\mathcal{O} (m^4 z^4)\,, \nonumber \\
\chi(z)&=&\frac{m}{2}\,z + \frac{m^3}{6}z^3\left (\log (\frac{m z}{2})+\frac14  \right )
+ \mathcal{O} (m^4 z^4) \,,\label{gulp6} \\
A(z)&=&-\frac{1}{12} m^2 z^2 + \mathcal{O} (m^4 z^4)\,,
\qquad B(z)=0\,. \nonumber
\end{eqnarray}

Turning now to the entanglement entropy, our goal is to compare our holographic
results to those found at weak coupling. To facilitate this comparison, we must
perform the holographic calculation for a `waveguide' geometry analogous to
those considered in the previous section. Given the above ansatz \reef{anzata},
the boundary geometry is just flat space, but if two of the spatial coordinates
are periodic, this geometry can be interpreted as a toroidal waveguide
$\BR^2\times \BT^2$. That is, we will identify the spatial coordinates, $x^1$
and $x^2$, with a period $H$ where $H\gg 1/m_b,\,1/m_f$. The entangling surface
$\Sigma$ is then chosen as $x^3=0$ (and $t=0$). According to eq.~\reef{define},
we must find the extremal bulk surface that connects to this entangling surface
as $z\to 0$. However, the high degree of symmetry here dictates that the
extremal surface will simply fall straight into the bulk geometry. That is, the
desired bulk surface is simply given by $\m=\lbrace x^3=0\,,t=0\rbrace$. Next
we must evaluate the area of this surface, however, as usual, the latter must
be regulated by cutting off the integral at the UV regulator surface
$z=\delta$. We are particularly interested in identifying any contributions
proportional to $\log\delta$ and so it is sufficient to consider the asymptotic
solution \reef{azymp} in calculating the area.
\begin{eqnarray}
 A(\m)&=&\int dx^1 dx^2 dz\, \sqrt{g_{11} g_{22} g_{zz}}
 =H^2 \int_{\delta} dz\, \frac{L^3}{z^3}\, e^{2 A(z)}\nonumber\\
&\simeq
& H^2L^3\int_{\delta} \frac{dz}{z^3}\,\left(
1-\frac{2}{3}\,\chi_{0,0}^2\,z^2+ O(z^4 \log^2 z)\right)
\label{house}\\
&\simeq& \frac{H^2 L^3 }{2 \delta^2}+\frac{2}{3}H^2 L^3\, \chi_{0,0}^2 \log
\delta+ \mathcal{O}(\delta^2\log^2\!\delta)\,.  \nonumber
 \end{eqnarray}
Now we may combine this result with eqs.~\reef{g5} and \reef{hum3} to write the
entanglement entropy as
 \be
S_\mt{EE}  =   \dfrac{A(\m)}{4 G_{5}} \simeq \dfrac{N^2 \mathcal{A}_{\Sigma}}{4
\pi \delta^2}+\dfrac{N^2 \mathcal{A}_{\Sigma}}{12 \pi} m_f^2 \log(m_f\delta)+
\cdots \label{boom0}
 \ee
where $\mathcal{A}_{\Sigma}=H^2$ is the area of the entangling surface in the
boundary theory. Of course, the leading contribution here is the expected `area
law' contribution, which does not yield any universal information.  The next
term yields the desired logarithmic term, with the same general form
\reef{franks} as found from free field calculations. We may observe that, as is
already evident in eq.~\reef{house}, neither of these UV divergent terms
depends on the higher order coefficients in the asymptotic expansion
\reef{azymp}. Hence as expected \cite{relevant}, our calculation explicitly
shows that these contributions to $S_\mt{EE}$ are insensitive to the IR details
of the holographic RG flow. Of course, we may also observe that only the
fermionic mass $m_f$ appears in the logarithmic term and the result does not
depend on the bosonic mass $m_b$.

While the above result applies to the gauge theory at strong coupling, we can
calculate the analogous contribution to the entanglement entropy in the weak
coupling limit. In this case, we simply consider the free field limit and apply
the results given in eqs.~\reef{coeff} and \reef{coeff2} in eq.~\reef{franks}.
If we first turn to $\calo_2$ in eq.~\reef{massb}, we see that there are three
complex scalars which acquire masses. Since each of these fields is in the
adjoint representation of the $U(N)$ gauge group, each $\phi_i$ effectively
contributes $2N^2$ real scalars in the free field limit. Hence using
eq.~\reef{coeff}, the total coefficient in the logarithmic contribution becomes
 \be
\gamma_{4,b} =\frac1{24\pi}\, \sum 2N^2\,m^2_i
=\frac{N^2}{12\pi}\left(\frac{m_b^2}3+\frac{m_b^2}3 -\frac{2m_b^2}3\right)
=0\,.
 \label{gammab}
 \ee
Therefore in the weak coupling limit, the bosonic mass term does not in fact
contribute to this universal area term in the entanglement entropy. Remarkably,
this agrees with our strong coupling result \reef{boom0}, in that the bosonic
mass $m_b$ does not appear in this expression.

Next, we consider the contribution of $\calo_3$ in the free field limit. In
this case, all three scalars again acquire masses and the two fermions
$\psi_{1}$ and $\psi_2$ also acquire a mass. In the massless limit, $\psi_1$
and $\psi_2$ are two independent Weyl fermions but here they combine as a
single massive Dirac fermion. Again the latter are in the adjoint
representation and so effectively we have $N^2$ Dirac fermions in the free
field limit. Hence combining the results in eqs.~\reef{coeff} and
\reef{coeff2}, the total coefficient in the logarithmic contribution becomes
 \be
\gamma_{4,f}=\frac1{12\pi}\,N^2\,m_f^2 +\frac1{24\pi}\,2N^2\,
\left(\frac{2m_f^2}3+\frac{2m_f^2}3 +\frac{2m_f^2}3\right) =
\frac{N^2}{4\pi}\,m_f^2\,. \label{gammaf}
 \ee
Comparing the above with the strong coupling result \reef{boom0}, we see that
this coefficient is larger by a factor of three than that appearing at strong
coupling. We can combine eqs.~\reef{gammab} and \reef{gammaf} to evaluate the
coefficient in the supersymmetric theory with $m_f=m_b=m$:
 \be
\gamma_{4,susy}=\gamma_{4,f}(m_f=m) + \gamma_{4,s}(m_b=m)  =
\frac{N^2}{4\pi}\,m^2\,. \label{gammasusy}
 \ee
Similarly, this choice of masses does not effect the holographic result
\reef{boom0} and so the discrepancy observed above between the coefficients in
the strong and weak coupling limits extends to the supersymmetric case.

Note that the strong coupling coefficient seems to match the contribution of
the fermions alone at weak coupling. That is, we would have found agreement
between the two limits if the `fermionic' mass term $\calo_3$ only gave mass to
the fermions $\psi_1$ and $\psi_2$. However, as is evident from
eq.~\reef{massf}, this dimension-three operator also contains a mass term for
all three scalars. The latter is somewhat unusual as it is a
`coupling-dependent' correction induced at finite mass, \ie $m_f\,\calo_3$
contains a contribution of order $m_f^2$. However, the presence of the
additional interactions in eq.~\reef{massf} is dictated by the supersymmetry
algebra and the global $SO(6)_R$ symmetry of the $\caln=4$ theory \cite{opera}.
Further, one can directly detect these scalar masses \cite{cheque} by uplifting
the asymptotic solution \reef{azymp} to ten dimensions \cite{n22} and then
examining the potential felt by a probe D3-brane, \eg \cite{bpp}. Of course, we
should also note that beyond the fermion and scalar mass terms, $\calo_3$ also
contains trilinear couplings between the hypermultiplet scalars $\phi_{1,2}$
and the gauge multiplet scalar $\phi_3$. Hence $m_f$ does not simply
parameterize the masses of various fields but also plays a role as the coupling
of a new cubic potential term in the interacting theory. The latter may well be
the source of the discrepancy observed above.

\section{Discussion} \label{discuss}

In this paper, we studied the entanglement entropy entropy for a variety of
field theories on waveguide geometries with a spherical or a hyperbolic
cross-section. Our calculations confirmed the appearance of a universal area
term \reef{franks}, which was first identified in \cite{wilczek}, as well as
reproducing the precise coefficient \reef{coeff} for a free massive scalar
field. Our analysis of the scalar fields in section \ref{scalar} also included
a nonminimal curvature coupling $\frac12 \xi R\, \phi^2$ and we found the
previous area term is not affected by this new coupling. In section
\ref{fermion}, we also considered a free massive fermion and we identified the
same universal area terms as in eq.~\reef{franks} with the coefficient given in
eq.~\reef{coeff2}. This reproduces a result given previously in \cite{new}.

\subsection*{Curvature contributions:}

By considering waveguide geometries with a curved cross-section, we were also
able to consider new curvature contributions in the entanglement entropy of the
schematic form shown in eq.~\reef{univEE}. Such terms were first identified
with holographic techniques \cite{relevant}, however, we can now present a
clearer understanding of the origin of such contributions to entanglement
entropy, following the perspective given in, \eg \cite{hong9,berk,cusp}. As
noted in the introduction, the calculation of entanglement (or R\'enyi) entropy
in QFT generically yields a UV divergent answer because the result is dominated by the
short distance correlations in the vicinity of the entangling surface $\Sigma$.
Hence the calculation must be regulated by introducing a short distance cut-off
$\delta$, and the result typically contains a series of power law divergences,
as shown in eq.~\reef{diverg0} for a QFT in $D$ spacetime
dimensions.\footnote{Of course, with even $D$, the $c_D$ term may appear with a
logarithmic divergence.} Of course, the leading contribtuion has a geometric
structure, in that it corresponds to the famous `area law' term \cite{bh9} with:
 \be
 c_2=\int_\Sigma d^{D-2}\sigma\,\sqrt{\gamma}\ d_2 = d_2\ {\cal
 A}_\Sigma\,,
 \label{coeff0z}
 \ee
where $\gamma_{ab}$ is the induced metric on the entangling surface. However,
if we are working with a covariant regulator (in a relativistic QFT) and
assuming the short-distance cut-off is much smaller than any scale defined by
the couplings of the QFT, \ie $\delta\mu_i\ll1$, then in fact all of the
coefficients of divergent terms in this expansion \reef{diverg0} exhibit a
similar geometric structure. For example, the second coefficient may be written
as
 \bea
c_4&=&\int_\Sigma d^{D-2}\sigma\,\sqrt{\gamma}\, \left[\,d_{4,1}\, {\cal
R}(\gamma)+ d_{4,2}\,R^{ij}\,\tilde{g}^\perp_{ij}+ d_{4,3}\,R^{ijkl}\,
\tilde{g}^\perp_{ik}\, \tilde{g}^\perp_{jl}\right.
  \label{coeff2z}\\
&&\qquad\qquad\qquad\qquad\qquad\left. + d_{4,4}\, K^\hi_a{}^b K^\hi_b{}^a+
d_{4,5}\, K^\hi_a{}^a K^\hi_b{}^{\,b}\,\right]\,,
 \nonumber
 \eea
where, \eg $\cal R(\gamma)$ denotes to the intrinsic Ricci scalar of the
entangling surface.\footnote{We refer the interested reader to \cite{EEGB} for
a full explanation of the notation.} This geometric character of the
coefficients naturally follows from the fact that the UV divergences are all
local.

The dimensionless coefficients $d_{2k,a}$ above will of course depend on the
detailed structure of the underlying QFT. However, because the $d_{2k,a}$ are
dimensionless, we can write their dependence on any mass scale $\mu_i$ in the
QFT in terms of the dimensionless combination $\mu_i\delta$, \ie $d_{2k,a} =
d_{2k,a}(\mu_i\delta)$. Unfortunately, the coefficients appearing in the
expansion above are scheme dependent. Clearly, if we shift $\delta \to
\alpha\delta$, we find $d_{2k,a} \to \hat{d}_{2k,a}= \alpha^{2k-D}
d_{2k,a}(\alpha \mu_i\delta)$. Hence the regulator dependence here comes both
from the implicit dependence on mass scales in the QFT and the `classical'
engineering dimension of the individual coefficients $c_{2k}$. Of course, the
latter reasoning can be evaded in certain special circumstances. One well-known
example, which was already noted in the introduction is the case of a CFT in an
even number of spacetime dimensions \cite{finn,adam,rt2,solo,cthem2}. Given the
underlying field theory is conformal, there are no intrinsic mass scales and in
an even spacetime dimension, the $c_{D}$ term still accompanies a logarithmic
divergence. Hence the corresponding coefficients $d_{D,a}$ evade the above
scaling argument and provide universal information about the underlying field
theory. In fact, as illustrated in eq.~\reef{WaldS9}, these coefficients are
proportional to the various central charges of the CFT.

Another example, which we might consider, is where the underlying field theory
describes some renormalization group flow beginning at a UV fixed point. The
theory that is probed by the UV singularities \reef{diverg0} in the
entanglement entropy is a CFT perturbed by some relevant operators. Now let us
consider the special case where the RG flow in the UV is controlled by a
single\footnote{With more than one scale, one could carry out the following
analysis with the largest scale $\mu_0$ while allowing the coefficients
$d^{(n)}_{2k,a}$ to be functions of the ratios $\mu_i/\mu_0$.} mass scale
$\mu$. By assumption $\mu\delta\ll1$ and so we can express the coefficients in
terms of a Taylor series: $d_{2k,a}=\sum_{n=0} d^{(n)}_{2k,a}\, (\mu\delta)^n$.
At this point, it is straightforward to extend the previous discussion to show
when the coefficient $d^{(n)}_{2k,a}$ with $n=D-2k$ does not scale with
$\alpha$. Hence, we can expect that this coefficient provides universal
information about the underlying RG flow. Ultimately, we may repackage this
discussion to see that we have identified possible universal contributions to
the entanglement entropy which take the form
 \be
S_\mt{EE}\simeq d^{(D-2k)}_{2k,a}\, \mu^{D-2k} \int_\Sigma d^{D-2}\sigma
\,\sqrt{\gamma}\, \left[(``curvature")^{k-1}\right]\,,
 \label{diverg3z}
 \ee
which, of course, matches the form of the expressions given previously in
eq.~\reef{univEE}.

Of course, the previous discussion can not be complete as we know that
universal contributions can also appear with logarithmic factors
$\log(\mu\delta)$, as illustrated in eqs.~\reef{franks} and \reef{newX}. So in
general, we must allow for such $\log(\mu\delta)$ factors in the expansion of
$d_{2k,a}$ to account for these situations. At this point, we would like to
note that holographic techniques were used to study the effect on entanglement
entropy of perturbing a strongly coupled CFT with relevant operators in
\cite{relevant}. One of the interesting results there was to show that
logarithmic contributions can appear in either even or odd spacetime
dimensions, in situations where the operators had large anomalous dimensions.
For example, perturbing by an operator with conformal dimension
$\Delta=(D+2)/2$ would generate such a $\log\delta$ contribution in general.
Further the holographic calculations in \cite{relevant} show that such large
anomalous dimensions may also introduce unusual powers in the expansion of the
coefficients $d_{2k,a}$.

The appearance of universal contributions as in eq.~\reef{univEE} were first
uncovered with a holographic approach in \cite{relevant} and with sufficient
effort, these holographic calculations allowed their precise geometric form to
be identified in a straightforward way. However, this identification is much
more difficult in the context of field theory calculations presented in this
paper.  In fact, the geometric nature of various universal contributions is
easily confirmed as follows. First, we observe that for sufficiently large
masses, our results are naturally be presented in terms of an expansion in
powers of $1/(mR)^2$, as could be anticipated from previous calculations
\cite{new,igor9,klebnew}. Of course, this is precisely in agreement with the
geometric expansion discussed above or given in eq.~\reef{univEE}, where the
powers of $1/R^{2n}$ correspond to factors of $[``curvature"]^n$. This
identification is further confirmed by repeating the calculations for
hyperbolic waveguides in Appendix \ref{renyiSd}. There we found that the same
expansion is produced up to the replacement $1/(mR)^2 \to -1/(mR)^2$, as
compared to the spherical waveguides in section \ref{wave}. Of course, this
sign corresponds precisely to the change in the sign of the curvature between
these two families of geometries.

While in general extracting the precise geometric form of these contributions
is difficult with the heat kernel approach used here, the simplicity of our
background geometries allowed us to identify the first such universal
contribution. As shown in eq.~\reef{newX}, it is simply an integral over the
entangling surface of the Ricci scalar evaluated on this surface. In general,
one expects that five independent curvature terms would contribute at this
order, as shown in eq.~\reef{coeff2z}. However, we have chosen a particular
`basis' for these terms there such that the expression identified in
eq.~\reef{newX} is the only nonvanishing term for the waveguide geometries
studied here. To be precise, the last two possible contributions in
eq.~\reef{coeff2z} involve an integral of terms quadratic in the extrinsic
curvature of the entangling surface. However in the present construction, the
extrinsic curvature is precisely zero and so these contributions vanish here.
Similarly, the second and third contributions in eq.~\reef{coeff2z} involve the
background curvature in the space transverse to the entangling surface. In the
present case, this transverse geometry is simply $\BR^2$ and so these
contributions are also zero.

The coefficient of the curvature contribution identified in eq.~\reef{newX} is
given in eqs.~\reef{coeff9x} and \reef{coeff8x} for the free scalar and fermion
fields, respectively. In the scalar case \reef{coeff9x}, we see that this
coefficient depends on the non-minimal coupling $\xi$. Focussing on this linear
$\xi$ dependence, we note that the coefficient is precisely that which would
appear if we replaced the mass $m$ in the corresponding area term \reef{franks}
by the effective mass $m_0$ appearing in the action \reef{scalaract}, \ie
 \be
m_0^{D-2}= \left(m^2+\xi\, {\cal
R}\,\right)^{(D-2)/2}=m^{D-2}+\frac{D-2}{2}m^{D-4}\,\xi\, {\cal R} +\cdots \
.\label{cute}
 \ee
Here we should keep in mind that these results, \ie eqs.~\reef{firstR} and
\reef{firstRx}, were derived from expressions where we had assumed $mR\gg1$.
Hence it is reasonable to consider the Taylor series expansion above. It seems
that this simple dependence originates with the appearance of $m_0^2$ in the
definition \reef{deff} of the parameter $b^2$ which characterizes the mass
dependence of our heat kernels. We might also note that $D=4$ seems
distinguished in eq.~\reef{coeff9x} or \reef{firstRx} in that this contribution
vanishes when $\xi$ takes the value for a conformal scalar.

As an aside, we point out here that there are finite contributions to the
entanglement (and R\'enyi) entropy which do not fall into the class of terms
discussed above. For example, as shown in eqs.~\reef{simp1} or \reef{simp3},
our results for the spherical waveguides revealed certain `topological'
contributions which are exponentially suppressed. We referred to these terms as
topological because there were no analogous contributions for waveguides with a
hyperbolic cross-section. However, we also see in, \eg eq.~\reef{simp3}, that there are contributions with inverse powers of the
mass. Recently, it has been shown that such terms with negative powers of $m$
also have a universal character, as they can be related to the universal
contributions appearing in entanglement entropy in higher dimensions
\cite{new,igor9}. Note that if we examine the universal $\log\delta$
contribution in even $D$, none of the individual terms in this coefficient have
the topological character or inverse powers of the mass described here. The
same is true for the coefficients $c_{2k}$ of the power law divergences in
eq.~\reef{diverg0}.

\subsection*{\ren entropy:}

Our calculations also extend the initial work of \cite{wilczek} by presenting
expressions for the \ren entropy, as well as the entanglement entropy. In
general, our results for the \ren entropy have the same structure as the
entanglement entropy. In fact, for any of the geometries or field theories
which we studied here, we can write
 \be
 S_\alpha=\frac{1+\alpha}{2\alpha}\,S_\mt{EE}\,.
 \label{widget}
 \ee
Hence in general, there are a variety of universal contributions in $S_\alpha$
which take the same geometric form as described above or in eq.~\reef{univEE}.
Unfortunately, with the relation in eq.~\reef{widget}, the \ren entropy would
not provide any new information that is not already available in the
entanglement entropy. However, it seems that this simple `factorization' of the
\ren entropy must be an artifact of the simplicity of both the background
geometries and the QFT's studied here. Typically, the \ren entropy has a more
complex dependence on the index $\alpha$ than appears in eq.~\reef{widget} as
can be seen, \eg in the holographic results of \cite{ren} or the results for
disjoint intervals in \cite{tonni}. Both of these examples also demonstrate
that the \ren entropy often contains far more information about the underlying
field theory than the entanglement entropy alone.

The form of the \ren entropy in eq.~\reef{widget} is reminiscent of well-known
results derived in two dimensions \cite{cardy0}. We can compare our expressions
for the \ren entropy for $D=2$ ($d=0$) to those given in \cite{cardy0} as a
check of our calculations. In that reference, the authors consider a
two-dimensional CFT perturbed by a relevant operator which introduces a
correlation length $1/m$. Combining various expressions appearing there, the
\ren entropy becomes
\begin{equation}
S=-\frac{1+\alpha}{12\,\alpha}\,c\,\log(m\delta)\,,
\label{scalarr}
\end{equation}
where $c$ is the central charge of the CFT defining the UV fixed point. In our
approach, the result for a massive scalar can be derived by substituting
eq.~\reef{sintKc2} directly into eq.~\reef{spartition} and the answer then
takes precisely the form given above with $c=1$ as is appropriate for a free
scalar field (in the conventions of \cite{cardy0}). Similarly substituting
eq.~\reef{intKc2fer2} into eq.~\reef{spartition} yields a result for a massive
Dirac fermion, which again has the form given in eq.~\reef{scalarr} with $c=1$.
Here, we might note that eq.~\reef{spartition} has an extra minus sign in the
case of the fermion relative to the scalar. However, eq.~\reef{intKc2fer2} also
has an extra minus sign in comparison to its scalar counterpart and so we
obtained the desired match.

\subsection*{Strong coupling:}

Previously, ref.~\cite{relevant} used holographic techniques to study the
effect on entanglement entropy of perturbing a strongly coupled CFT with
relevant operators. In the gravity description, such an operator is dual to a
scalar field and the holographic entanglement entropy \reef{define} is modified
by the backreaction of the scalar on the geometry through Einstein's equations.
As noted above, one of the interesting results was that logarithmic
contributions could appear in either even or odd spacetime dimensions, in
situations where the operators had large anomalous dimensions. For example,
perturbing by an operator with conformal dimension $\Delta=(D+2)/2$ would
generate such a $\log\delta$ contribution in general. Another result, which
seemed to present a small puzzle, was that an operator with the dimension of a
scalar mass term, \ie $\Delta=D-2$, generated a $\log\delta$ term in even
dimensions but only for $D\ge6$. The puzzle was then that at weak coupling, \ie
with free field theories, such a logarithmic term appears for a massive scalar
\cite{wilczek}, as shown in eq.~\reef{franks}.

Here in section \ref{strong}, we considered a particular holographic framework
where the AdS/CFT dictionary is very well understood, namely the
four-dimensional $\caln=2^*$ gauge theory. In this case, two mass operators
with $\Delta=2$ and 3, as well as their holographic description in the dual
gravity theory, are known exactly. The detailed description of the bosonic mass
term \reef{massb} in the boundary theory seems to resolve the puzzle noted
above. In particular, the intuition provided by weak coupling is that the
coefficient of the $\log\delta$ term is proportional to $\sum m_i^2$ in a case
where several scalars acquire masses. However, this sum is precisely zero for
the present mass term given in eq.~\reef{massb}. Hence rather than a puzzle, we
have precise agreement between the weak and strong coupling results in this
specific case.

In examining the effect of the `fermionic' mass operator \reef{massf} on the
entanglement entropy, we found a contribution in eq.~\reef{boom0} of the
general form $\mathcal{A}_\Sigma m_f^2\log(m_f\delta)$, as expected an
analogous weak coupling calculation. Unfortunately, if we compare the precise
coefficient found at strong coupling with that appearing in the free field
limit, there is a discrepancy by a factor of three. Note that the latter adds
contributions from both the fermions and scalars which acquire masses when this
operator is introduced. As shown in eq.~\reef{massf}, $\calo_3$ also contains a
cubic interaction between the scalar fields, which vanishes in the weak
coupling limit, \ie $g_\mt{YM}\to0$. However, in the present context then,
$m_f$ does not only parameterize the masses of various fields but it also plays
a role as the coupling of a new cubic potential term in the interacting theory.
Of course, it is tempting to argue that the latter is the source of the
discrepancy observed between the strong and weak coupling results. An
interesting extension of the present results then would be to calculate the
effect of the cubic interaction on the entanglement entropy perturbatively when
the gauge coupling is small but finite.

We would like to contrast the above discrepancy with the recent results in
\cite{hertz}. There the author found in perturbative calculations, that the
effect of the interactions on the entanglement entropy was to properly
renormalize the mass appearing in eq.~\reef{franks} so that it corresponded to
the physical mass. Beyond the usual difficulties that one would encounter in
extending such a calculation to strong coupling, our holographic calculation
points out another difficulty in taking this limit. Namely, our strongly
coupled boundary theory has no simple (quasi)particle excitations and hence we
could not identify the `renormalized mass' from a pole in a two-point function.
Clearly, a better understanding is needed to appreciate the precise sense in
which the entanglement entropy contributions identified in \cite{wilczek} are
universal or alternatively, to unravel the precise information that these terms
carry about the underlying field theory. It would be useful study further
holographic examples where the precise definition of the relevant operators is
known in both the bulk gravity and boundary field theories. Another example,
which would be interesting for this purpose, is the Cvetic-Gibbons-Lu-Pope
solution \cite{cglp}, which describes an RG flow from a three-dimensional CFT
\cite{cft3}) in the UV to a gapped theory in the IR.

\subsection*{Comparison with CFT's:}

As noted previously, certain universal contributions to the entanglement
entropy are well known in the case of CFT's in an even number of spacetime
dimensions \cite{finn,adam,rt2,solo,cthem2}. While our calculations focussed on
the universal contributions appearing with masses, they should also yield the
expected CFT results in the appropriate limits. Hence it is interesting to
compare our expressions with the expected CFT results as a check of our
calculations.

The universal contribution for a four-dimensional CFT \cite{solo} is given in
eq.~\reef{WaldS9}. Setting $D=4$ in section \ref{wave}, the corresponding
entangling surface is $\Sigma=\BS^2$. In our four-dimensional waveguide
geometry, the extrinsic curvatures vanish, $\cR_\Sigma=\cR_{\BS^2}=2/R^2$ and
$C^{\mu\nu\rho\sigma}\,h_{\mu\rho} \,h_{\nu\sigma}=\cR_{\BS^2}/3=2/(3R^2)$.
Hence, eq.~\reef{WaldS9} yields
 \be
S_\mt{univ}=4\,\( a-\frac{c}3 \)\, \log(\delta/R)~.
 \label{resultx9}
 \ee
Moreover, the central charges for a massless conformal scalar ($\xi=1/6$) and a
massless fermion are \cite{Birrell}:
 \be
 a=\begin{cases}
  {1\over 360} & \text{for $s = 0$}\,, \\
  {11\over 360} & \text{for $s=1/2$}\,,
\end{cases}
\quad{\rm and}\quad
 c=\begin{cases}
  {1\over 120} & \text{for $s = 0$}\,, \\
  {1\over 20} & \text{for $s=1/2$}\,.
\end{cases}
 \label{charges}
 \ee
Therefore, we finally find
 \be
S_\mt{univ}(s=0)=0\,, \quad{\rm and}\quad S_\mt{univ}(s=1/2)=\frac1{18}\,
\log(\delta/R)\,.
 \label{resultc}
 \ee
Now we may compare these results with those derived in section \ref{wave}. In
particular, if we set $m=0$ and $\xi=1/6$ (and $\al=1$) in eq.~\reef{sren4dS},
we see that the $\log\delta$ term in the entanglement entropy vanishes for a
conformal scalar, in agreement with the above result. Similarly if we set $m=0$
(and $\al=1$) in eq.~\reef{ferren4dS}, we recover precisely the above
expression for the universal contribution of a massless Dirac fermion.

Our waveguide geometry also lends itself to using the approach of \cite{cthem2}
to determine the universal contribution for a CFT in any number of dimensions.
The only restriction of this latter approach is that the background geometry
must have a rotational symmetry in the transverse space around the entangling
surface, which is certainly satisfied in the present case. The result for the
six-dimensional waveguide $\BR^2\times \BS^4$ is given in \cite{EEGB} as
 \be
S_1={9\pi \cA_{\Sigma} \over 50 R^4}\( {25\over 3\pi^3}A-17B_1+52B_2-592B_3 \)
\log(m\delta) \label{sixD}
 \ee
where $A,\ B_1,\ B_2,$ and $B_3$ are the four central charges which appear in
the trace anomaly of a general CFT in $D=6$. These central charges were found
in \cite{anom} for a massless conformal
scalar ($\xi=1/5$):
 \be
A=-{5\over 3\cdot 7!}\, ,~ B_1={28\over 3(4\pi)^3\,7!}\,,~B_2=-{5\over
3(4\pi)^3\,7!}\,,\ \ {\rm and}\ \ B_3=-{2\over (4\pi)^3\,7!}~;
 \label{censcal}
 \ee
and for a massles Dirac fermion:
 \be
A=-{191\over 3\cdot 7!}\, ,~ B_1={896\over 3(4\pi)^3\,7!}\,,~B_2={32\over
(4\pi)^3\,7!}\,, \ \ {\rm and}\ \ B_3=-{40\over (4\pi)^3\,7!}~.
\label{cenferm}
 \ee
Thus the universal contribution \reef{sixD} to the entanglement entropy becomes
 \be
 S_\mt{univ}=\begin{cases}
\frac{\pi}{67500} \, \log(m\delta) & \text{for a massless conformal scalar}\,, \\
-\frac{11\pi}{2700} \, \log(m\delta) & \text{for a massless Dirac fermion}\,.
\end{cases}
 \label{sixDz}
 \ee
To extract the corresponding expression for the fermion from our results, we
consider eq.~\reef{6Dferm} for the universal contribution on a hyperbolic
waveguide in $D=6$. Replacing Vol$(\BH^4)$ with Vol$(\BS^4)=8\pi^3R^4/15$ and
$1/R^2\to -1/R^2$ in the subsequent factors, we have the analogous contribution
for a spherical waveguide. Then setting $m=0$ and $\alpha=1$, we recover
precisely the universal contribution for a massless fermion given above in
eq.~\reef{sixDz}. The corresponding expression for the conformal scalar can be
obtained directly from eq.~\reef{6Dscal} after setting $m=0$, $\xi=1/5$ and
$\alpha=1$. Unfortunately the result in this case is: $S_\mt{univ} =-
\frac{\pi}{13500} \, \log(m\delta)$. Hence we have a discrepancy of a factor of
--5 between our calculation and the corresponding result in eq.~\reef{sixDz}!
Unfortunately, at this point, we have not been able to track down the source of
this discrepancy despite checking both approaches in many ways, \eg verifying
the central charges in eq.~\reef{censcal} with the heat kernel techniques used
here.

\acknowledgments

We would like to thank Alex Buchel, Horacio Casini, Ling-Yan Hung and Ben Safdi
for useful conversations. AL would like to acknowledge support from Fundacion
Caja Madrid and the Perimeter Scholars International program. Research at
Perimeter Institute is supported by the Government of Canada through Industry
Canada and by the Province of Ontario through the Ministry of Research \&
Innovation. RCM also acknowledges support from an NSERC Discovery grant and
funding from the Canadian Institute for Advanced Research.

\appendix

\section{Separation of variables in the case of spin-${1\over 2}$ fields}
\label{appx:sepvar}

The main goal of this appendix is to prove the separation of variables
\reef{sepheat} in the case of spin-${1\over 2}$ fields. Here we focus on the
waveguide geometry $\cM_\al=C_\al\times\BS^d$ discussed throughout the main
text. However, in Appendix \ref{renyiSd}, we also consider the hyperbolic
waveguide $C_\al\times\BH^d$ and hence we note that the present discussion is
equally well applicable to the latter case since the essential ingredient is
the product structure of the waveguide geometry. We begin by reviewing our
spinor notation. The spinors are associated with the orthonormal frames
$e^{\mu}_a$ on $\cM_\al$
 \be
 e_{a}^{\mu}\ e_{b}^{\nu}\ g_{\mu\nu}=\delta_{ab}~,
 \ee
whereas the Clifford algebra in the orthonormal frame is generated by
$D$ matrices $\gamma^a$, satisfying the anticommutation relations
 \begin{equation}
 \{\gamma^a,\gamma^b\}=2\,\delta^{ab}~.
 \label{clifford}
 \end{equation}
The dimension of these matrices is $2^{\lfloor{D\over 2}\rfloor}$ and
the associated $D(D-1)/2$ generators of $SO(D)$ rotations are given by
 \be
 \sigma^{ab}={1\over 4}[\gamma^a,\gamma^b]~.\label{lorentz}
 \ee
The latter satisfy the standard $SO(D)$ commutation rules
 \be
[\sigma^{ab},\sigma^{cd}]=\delta^{bc}\sigma^{ad}-
\delta^{ac}\sigma^{bd}-\delta^{bd}\sigma^{ac}+ \delta^{ad}\sigma^{bc}~,
 \label{comrul}
 \ee
while the commutator of $\sigma^{ab}$ with $\gamma^c$ is
 \be
 [\sigma^{ab},\gamma^c]=\delta^{bc}\gamma^a-\delta^{ac}\gamma^b~.
 \label{comrul2}
 \ee

The covariant derivative of a spinor may be written in terms of
$e^{\mu}_a$ as follows:
 \be
 \nabla_a=e^{\mu}_a\nabla_{\mu}~, \quad
 \nabla_{\mu}=\partial_{\mu}+{1\over 2} \sigma^{bc} \omega_{\mu b c}~, \quad
 \omega_{\mu b c}=e^{\nu}_b(\del_\mu e_{c\nu}-\Gamma_{\nu\mu}^\alpha e_{c\alpha})~,
 \label{spinorcov}
 \ee
where $\Gamma_{\nu\mu}^\alpha$ are the usual Christoffel symbols. With
this definition, the following anticommutation relations can be shown
to hold \cite{DeWitt}
 \begin{equation}
 [\nabla_a,\nabla_b]\,\psi=-{1\over 2}\,R_{abcd}\,\sigma^{cd}\,\psi~.
 \label{Dirac-comm}
 \end{equation}
In particular, since the $\gamma^{a}$ matrices are covariantly constant, one
can use eqs.~\reef{clifford} and \reef{Dirac-comm} to verify the following
identity for the iterated Dirac operator:
 \be
\slashed\nabla\cdot\slashed\nabla^{\dag}=-
\slashed\nabla^2=-(\gamma^a\nabla_a)^2=
-\delta^{ab}\nabla_a\nabla_b+{R\over 4}~.
 \label{nabla2}
 \ee
It is the heat kernel of this operator on $\cM_\al$ that we use in
eq.~\reef{spartition} to evaluate the partition function in the case of
spin-${1\over 2}$ fields.

Since the waveguide geometry has a product structure, it is natural to choose
the first two $\gamma$-matrices $a=1,2$ to construct the Dirac operator on
$C_{\al}$ and then the rest are used to build the restriction of Dirac operator
to $\BS^{d}$. Now, we observe that due to the various relations in
eqs.~\reef{clifford}, \reef{comrul} and \reef{comrul2}, the restriction of
Dirac operator to $C_\al$ anticommutes with the restriction of Dirac operator
to $\BS^{d}$. Therefore we can split the iterated Dirac operator into separate
wave operators on $C_\al$ and $\BS^{d}$.

It is convenient to make this discussion more explicit by choosing the
following representation of the Dirac matrices on $\cM_\al$:
 \bea
 \gamma^1&=&\sigma^1\otimes \mathbb{I}_d \ ,
\qquad
 \gamma^2= \sigma^2\otimes \mathbb{I}_d \ ,
 \label{dgamma}\\
\gamma^a&=&\sigma^3\otimes \hat\gamma^{a-2}\quad {\rm for} \ a=3,\cdots,D\,.
 \nonumber
 \eea
where $\sigma^i$ are the standard Pauli matrices while $\hat\gamma^a$ and
$\mathbb{I}_d$ is the Dirac matrices and unit matrix of the Clifford algebra on
$\BS^{d}$. Hence the latter matrices have dimension $2^{\lfloor
d/2\rfloor}\times 2^{\lfloor d/2\rfloor}$. With this explicit representation,
the Dirac operator takes the form
 \be
\slashed\nabla_{\cM_\al}=\slashed\nabla_{C_\al}\otimes \mathbb{I}_d
+\sigma^3\otimes\slashed\nabla_{\BS^{d}}
 \label{dirac}
 \ee
where $\slashed\nabla_{C_\al}$ and $\slashed\nabla_{\BS^{d}}$ are respectively
the Dirac operators that would be constructed on $C_\al$ and $\BS^{d}$ alone.
Similarly, the iterated Dirac operator appearing in the heat kernel becomes
  \be
-\slashed\nabla^2_{\cM_\al}=-\slashed\nabla^2_{C_\al} \otimes \mathbb{I}_d +
\mathbb{I}_2 \otimes\left(-\slashed\nabla^2_{\BS^{d}}\right)\,.
 \label{waved}
 \ee
Hence it is clear that the heat kernel of the iterated massless Dirac operator
on the waveguide $C_\al\times \BS^{d}$ can be written as a product of the
individual heat kernels, as in eq.~\reef{sepheat}. Hence the desired separation
of variables is proved.

Further we note that the spinors on $\cM_\al$ have dimension $2^{\lfloor
D/2\rfloor}=2\,2^{\lfloor d/2\rfloor}$. With the above construction, we see
that the full trace over these spinor indices is properly accounted for by
calculating the two spinor heat kernels on $C_\alpha$ and $\BS^d$ separately
and then simply taking their product in eq.~\reef{traces}. That is, we can
treat the spinor trace on the right-hand side of eq.~\reef{traces} as though we
separately tracing over the spinor spaces of the two component manifolds.

While the desired result has been established, let us make a few more comments.
Recall that the construction of the spinor heat kernels $\BS^d$ given in
eqs.~\reef{fermheatSodd} and \reef{fermheatSeven} involve a spinor matrix $\hat
U$ which parallel propagates a spinor between the two points in the heat
kernel. According to \cite{camp}, this `propagator' $\hat U(x,y)$ on $\BS^d$
can be formally written as follows
 \be
\hat U(x,y)=P \exp {1\over 2}\int_x^y \omega_{\mu b c}(t)\,\hat\sigma^{bc}
v^{\mu}(t) dt ~, \label{propell}
 \ee
where $\hat\sigma^{ab}$ are the $SO(d)$ generators \reef{lorentz} for the
$d$-dimensional Clifford algebra, now constructed with the Dirac matrices
$\hat\gamma^a$. Above, the integration is along the shortest geodesic
connecting $x,y\in \BS^{d}$, $P$ is the path-ordering operator and
$v^\mu(t)=dx^\mu/dt$ is the tangent vector to the geodesic
$x^{\mu}(t)\in\BS^d$. With the representation in eq.~\reef{dgamma}, this
propagator becomes $U(x,y)=\mathbb{I}_2 \otimes\hat U(x,y)$. Hence, given
eq.~\reef{waved}, this again demonstrates that the restriction of the iterated
Dirac operator to $C_\al$ commutes with the heat kernels on $\BS^{d}$ given in
eqs.~\reef{fermheatSodd} and \reef{fermheatSeven}. This is, of course,
necessary for eq.~\reef{sepheat} to hold.

Finally, let us comment that demonstrating the separation of variables in
eq.~\reef{sepheat} here crucially depends on the fact that we are studying the
iterated Dirac operator \reef{iterate} rather than working with the Dirac
operator directly. In contrast, the eigenspinors of the $\slashed\nabla$ on
$C_\al\times S^d$ (or $H^d$) cannot be obtained easily from the lower
dimensional eigenspinors on the component manifolds $C_\al$ and $S^d$. While
one might construct the heat kernel on $\cM_\alpha$ using eigenspinors, clearly
such a construction would obscure the desired separation of variables and our
discussion above demonstrates that in fact this approach is not needed.

\section{Waveguides with hyperbolic cross-section}
\label{renyiSd}

Throughout the main text, we considered the waveguide geometry with a spherical
cross-section. In this appendix, $\BS^d$ is replaced by a $d$-dimensional
hyperbolic space $\BH^d$ and we show that \ren entropy is essentially the same
with an obvious sign change $R^2\rightarrow -R^2$ in various curvature
contributions.

In the following, we mimic our previous discussion of the case of a spherical
cross-section. In particular, massive scalars and fermions will be considered
separately, and within each of these, odd and even dimensions $d$ require
separate consideration. Further, both the simple regularization of the heat
kernel \reef{spartition} and the $\zeta$-function approach \reef{spartition3}
will be used to evaluate the \ren entropy. Recall that while the regularization
in eq.~\reef{spartition} reproduces the whole structure of the \ren entropy,
the $\zeta$-function method \reef{spartition3} eliminates the power law
divergences and retains only finite and logarithmically divergent terms. The
general remarks and formulae of section \ref{wave} do not require any
modifications apart from obvious replacements $\BS^{d}\rightarrow \BH^{d}$ and
$R^2\rightarrow -R^2$ and will not be repeated in the following discussion.
Instead we focus on the evaluation the heat kernel $K_{\BH^{d}}$ and
$\zeta_{\BH^d}^{(s)}$.

\subsection{\ren entropy for a massive scalar} \label{scalarH}

There is a vast literature which considers the scalar heat kernel on the
hyperbolic space $\BH^d$, \eg see \cite{scaheat}
 \bea
K_{\BH^{2n+1}}^{(0)}(t,x,y)&=& {1\over (4\pi t)^{1/2}} \left({-1\over 2\pi R^2
\sinh\rho}\frac{\del} {\del \rho}\right)^n
 e^{-{n^2 t\over R^2}-\frac{\rho^2 R^2}{4t}} ,
 \label{scalarheat} \\
K_{\BH^{2n+2}}^{(0)}(t,x,y)&=&  e^{-{(2n+1)^2\over 4R^2}t}\left({-1\over 2\pi
R^2 \sinh\rho}\frac{\del} {\del \rho}\right) ^n  f_{\BH^2}^{(0)}(\rho, t),
  \label{scalarheat2}
 \eea
where $n$ is or a positive integer; $\rho$ is the geodesic distance between $x$
and $y$ measured in units of $R$; and
 \be
f_{\BH^2}^{(0)}(\rho, t)={\sqrt{2} R\over (4\pi t)^{3/2}   } \int_\rho^\infty
{\tilde\rho \, e^{-{R^2\tilde\rho^2\over 4t} } \over
\sqrt{\cosh\tilde\rho-\cosh\rho}} \, d\tilde\rho
  ~.
 \label{2dscalarheat}
 \ee

The scalar $\zeta$-function was computed in \cite{camp2} where it was shown
that for odd $d\geq 3$
 \be
\zeta_{\BH^d}^{(0)}(z)={ R^{2z-d} b^{1-2z}\text{Vol}(\BH^{d})\over
(4\pi)^{d/2} \Gamma(d/2)} \sum_{k=0}^{(d-1)/2} g_{k,d}^{(0)}\,
  b^{2k} B(k+1/2,z-k-1/2)\,,
  \label{oddsczeta}
 \ee
where $B$ denotes the usual beta function with
$B(x,y)=\Gamma(x)\,\Gamma(y)/\Gamma(x+y)$ and
 \be
 b^2=R^2 m^2-\xi\, d(d-1)+\frac{(d-1)^2}{4}~.\label{deffH}
 \ee
Note that $b^2$ here in the case of $\BH^d$ is the same from its counterpart
\reef{deff} on $\BS^d$ with the replacement $R^2\rightarrow -R^2$. The
coefficients $g_{k,d}^{(0)}$ were introduced previously --- see
eqs.~\reef{defg} and \reef{useg}, as well as the surrounding discussion. For
even $d\geq 4$
 \be
\zeta_{\BH^d}^{(0)}(z)={ R^{2z-d}\text{Vol}(\BH^{d})\over (4\pi)^{d/2}
\Gamma(d/2)}
  \sum_{k=0}^{(d-2)/2} h_{k,d}^{(0)}
\Big[ b^{2k+2-2z} B(k+1,z-k-1)-4 \int_0^\infty {x^{2k+1}\over (e^{2\pi
x}+1)(x^2+b^2)^{z}}dx\Big]
  \label{sczeta}
  ~,
 \ee
where the coefficients $h_{k,d}^{(0)}$ are given in eqs.~\reef{defh} and
\reef{useh} (as well as the surrounding discussion). In both
eqs.~\reef{oddsczeta} and \reef{sczeta}, one should think of
$\text{Vol}(\BH^{d})$ as a formal regulated volume for the cross-section of the
waveguide geometry. While we express our results below in terms of this volume,
it may be more appropriate to think in terms of the \ren entropy density along
the entangling surface.

\subsubsection*{Odd dimensions}

Let us assume that $d=2n+1$ and take the limit of coincident points in
eq.~\reef{scalarheat}, then $K_{\BH^{d}}^{(0)}(t,x,x)$ takes the following
general form \cite{cahu}
\begin{equation}
K_{\BH^{2n+1}}^{(0)}(t,x,x)=\frac{P_{n-1}^{(0)}(t/R^2)}{(4 \pi t)^{n+1/2} }
\,e^{-{n^2 t\over R^2}}~.
\label{oddball}
\end{equation}
where from \reef{scalarheat} it follows that $P_{n-1}^{(0)}(x)$ is polynomial of degree $n-1$ with rational coefficients. For
$n=0$, $P_{-1}^{(0)}(x)=1$ while for $n>0$, $P_{n-1}^{(0)}(x)$ is polynomial of
degree $n-1$ with rational coefficients:
 \be
P_{n-1}^{(0)}(x)=\sum_{j=0}^{n-1} a_{j,n-1}^{(s)}x^j ~.
 \label{polynom}
 \ee
For example, let us write out the first few polynomials
 \bea
 P_0^{(0)}(x)&=& 1~,
 \nonumber \\
 P_1^{(0)}(x)&=&1+{2\over 3}x~,
  \nonumber \\
 P_2^{(0)}(x)&=&1+2x+{16\over 15}x^2~,
  \nonumber \\
 P_3^{(0)}(x)&=&1+4x+{28\over 5}x^2+{96\over 35}x^3~.
 \label{spoly}
 \eea
As one may surmise from these examples, $a_{0,n-1}^{(0)}\equiv1$ for $n\ge0$.
To simplify the following discussion, we also denote $a_{j,-1}^{(0)}=0$ for
$j>0$. We may evaluate the \ren entropy by combining the above heat kernels
\reef{oddball} with the usual expressions given in the main text (\ie
eqs.~\reef{renyi}, \reef{spartition}, \reef{traces} and \reef{sintKc2}), which
yields
 \bea
S_{\al}^{(0)}&=&{ 1+\al \over 24\,\al} \, \text{Vol}(\BH^{2n+1})
\int_{\delta^2}^{\infty} {dt\over t} \, \frac{P^{(0)}_{n-1}(t/R^2)}{(4 \pi
t)^{n+1/2} }\, e^{-{b^2 \over R^2}t }~. \label{scalaroddEE}
 \eea
We can easily evaluate the first few divergent terms in this expression as
 \be
S_{\al, div}^{(0)}={ 1+\al \over 12\,\al\, } {\text{Vol}(\BH^{d}) \over
(4\pi)^{d/2}}{1\over \delta^{d}} \( {1\over d}+
 \,{a_{1,{d-3\over 2}}-b^2\over (d-2)} {\delta^2\over R^2}+\ldots \)
 \label{sarealaw}
 \ee
Here we have used that in the vicinity of $t=0$, $P_{n-1}^{(0)}(t/R^2)\simeq
1$. As expected, we see that the leading divergence obeys the expected `area
law.' Continuing to higher orders in $t$, one can reconstruct all the UV
divergences. We note that the structure of these divergences matches those for
the case of $\BS^d$ up to an obvious sign flip in the background curvature
$R^2\to -R^2$, \eg one might compare the above expression with $n=1$ ($d=3$) to
eq.~\reef{ren3a}.

A simple approach to evaluate the finite terms in $S_{\al}^{(0)}$ is to proceed
in the spirit of dimensional regularization by setting $\delta=0$ but treating
$d$ as unspecified variable in the integration over $t$ in
eq.~\reef{scalaroddEE}. This yields
 \be
S_{\al, finite}^{(0)}= { 1+\al \over 24\,\al} \,{ \text{Vol}(\BH^{d}) \over
(4\pi R^2)^{d/2} }\sum_{j=0}^{d-3\over 2} a_{j,{d-3\over 2}}^{(0)} \, \Gamma
(j-d/2) \, b^{d-2j}~.
 \label{srenyiodd}
 \ee
This expression is also valid for $d=1$ if we omit the sum and substitute
$j=0$. In the limit $mR\gg1$, the leading behaviour in eq.~\reef{srenyiodd} is
given by
 \begin{multline}
S_{\al, finite}^{(0)}={ 1+\al \over 24\,\al} \,{  (-1)^{D-1\over 2}\pi\over
(4\pi)^{D-2\over 2} \Gamma(D/2)} \cA_\Sigma \Big(m^{D-2}+{D-2\over
2}{m^{D-4}\over R^2}
 \\
\times\ \Big[{(D-3)^2\over 4}-(D-2)(D-3)\xi -a_{1,{D-5\over
2}}^{(0)}\Big]+\ldots\Big)
 \label{renfinHodd}
 \end{multline}
where $\cA_\Sigma=\text{Vol}(\BH^d)$ and we have again used
$a_{0,(D-3)/2}^{(0)}\equiv1$. Further we have simplified the above result with
the following identity
 \be
 \Gamma(1-D/2)\Gamma(D/2)={\pi\over \sin({\pi D\over 2})}=(-1)^{D-1\over
 2}\pi
 \label{idnty}
 \ee
where the last equality applies because we are only considering odd $D$. If we
compare this result with eqs.~\reef{franks} and \reef{coeff}, we see that the
expression gives precisely the expected area term. We can also compare this
result with eq.~\reef{renfinSodd} for the spherical waveguide with odd $D$. In
this case, the second contribution above matches the corresponding term in
eq.~\reef{renfinSodd} with $R^2\to-R^2$ if
 \be
 a_{1,{D-5\over 2}}^{(0)}={(D-3)(D-5)\over 12}~, \label{splaxh}
 \ee
which matches the linear coefficients in the examples given in
eq.~\reef{spoly}. Of course, the $R^2\to-R^2$ behaviour is precisely that expected
of a curvature contribution, as given in eq.~\reef{firstR}.

Alternatively we can apply the $\zeta$-function method \reef{spartition3} to
derive $S_{\al, finite}^{(0)}$. According to eq.~\reef{oddsczeta},
$\zeta_{\BH^d}^{(0)}(0)=0$ and therefore, from eqs.~\reef{renyi} and
\reef{sintKc2}, we find
 \be
 S_{\al, finite}^{(0)}= {1+\al\over 24\al}\,\text{Vol}(\BH^d)
 \left. {d\zeta_{\BH^d}^{(0)} \over dz} \right|_{z=0} \label{whoha}
 \ee
Comparing this result to eq.~\reef{generalodd1}, we see that the differences
between the $\BH^d$ and $\BS^d$ cases are not accounted for by a change in the
sign of the curvature alone. While the first term in eq.~\reef{generalodd1}
matches that the (entire) result above with the replacement $R^2\to -R^2$,
there is also an extra integral contributing on the right hand side of
eq.~\reef{generalodd1}. Recall that this integral vanishes in the limit $m
R\to\infty$, however, for large but finite $mR$, it contributes exponentially
suppressed terms to $S_{\al, finite}^{(0)}$, \eg see \reef{simp3}. In the main
text, we speculated that these exponential contributions probe the topology of
the waveguide geometry. In this case, it would be natural that they vanish here
where the $\BH^d$ cross-section is topologically trivial. Lastly, we note that
we may evaluate eq.~\reef{whoha} using eq.~\reef{sczeta} and compare the result
with \reef{renfinHodd}. Agreement in this comparison again yields
eq.~\reef{splaxh}.

\subsubsection*{Even dimensions}

Below we evaluate the structure of all of UV divergences for the particular
case of $d=2$. Then using $\zeta$-function approach, we evaluate (only) the
universal contributions for any even $d\geq 4$.
\newline
\\
{\it $d=2$ ($D=4$):}
\\
\newline
In this case, we substitute $n=0$ in eq.~\reef{scalarheat2}and take the limit
of coincident points
 \bea
 K_{\BH^2}^{(0)}(t,x,x)&=& e^{-{t\over 4R^2}}
f_{\BH^2}^{(0)}(0, t)={ e^{-{t\over 4R^2}} \over 2\pi^{3\over 2}  R \sqrt{t}}\
\int_0^\infty {x \, e^{-x^2} \over \sinh \Big({\sqrt{t} \, x \over  R}\Big)} \,
dx  \\
&=&{ e^{-{t\over 4R^2}} \over 2\pi^{3\over 2}  t} \int_0^\infty dx \, e^{-x^2}
\Big(1-{t \over 6 R^2}\, x^2+ \ldots\Big) ={ e^{-{t\over 4R^2}} \over 4\pi t }
\Big( 1 - {t\over 12 R^2} + \ldots\Big) ~,
  \nonumber
 \eea
This result is then used as usual to evaluate the \ren entropy, yielding
 \bea
S_{\al}^{(0)}&=& {1 +\al \over 96 \pi\al}\,  \text{Vol}(\BH^{2})
\int_{\delta^2}^{\infty} {dt\over t^2} \, e^{-{t\over R^2}\big(1/4-2\xi+R^2
m^2\big)} \Big( 1 - {t\over 12 R^2} + \ldots\Big)
  \\
&=& {1 +\al \over 48 \pi\al}\, \text{Vol}(\BH^{2})\left[{1 \over2 \delta^2} +
\Big( m^2-{6\xi-1\over 3 R^2}\Big)\log(m\delta)+\ldots\right]\,.
\label{sren4d}
 \eea
Hence the leading divergence reveals the expected `area law' behaviour, while
with $\al$=1, the second term is precisely the result given in
eqs.~\reef{franks} and \reef{coeff} for $D=4$. Further, this result agrees with
the $\BS^2$ counterpart \reef{sren4dS} with the usual $R^2\to -R^2$
replacement.
\\
\newline
{\it General even $d\geq4$.}
\\
\newline
While we could determine the entire pattern of divergences in the \ren entropy
in higher even dimensions by extending the method considered above, here we
focus our attention on only the universal contributions. The latter are readily
evaluated using the $\zeta$-function approach \reef{spartition3} instead.

Evaluating scalar $\zeta$-function \reef{sczeta} at $z=0$ and
substituting the result into eqs.~\reef{renyi} and \reef{spartition3},
we find that universal contribution to the \ren entropy is given by
 \be
 S_{\al}^{(0)}={1+\al \over 12\al} { \text{Vol}(\BH^{d})\over (4\pi)^{d/2} \Gamma(d/2)R^{d}}\sum_{k=0}^{(d-2)/2}
h_{k,d}^{(0)}\[   4\int_0^\infty {x^{2k+1}\over e^{2\pi x}+1}dx -{(-b^2)^{k+1} \over k+1}\] \log(m \delta)
~.
\label{srenyieven}
 \ee
Expanding in the limit $mR>>1$, yields
 \be
 S_{\al, univ}^{(0)}={1+\al \over 12\al} { (-1)^{D\over 2}\cA_\Sigma\over (4\pi)^{D-2\over 2} \Gamma(D/2)}
 \Big(m^{D-2}-{(D-2)^2(D-3)(6\xi-1)\over 12}{m^{D-4}\over R^2}+\ldots\Big) \log(m \delta) \, ,
 \ee
where $\cA_\Sigma=\text{Vol}(\BH^d)$. The above results are in full agreement
with their counterparts for the spherical waveguide in eqs.~\reef{srenyievenS}
and \reef{exprenSev} provided that we replace $R^2\to -R^2$.

\subsection{\ren entropy for a massive fermion} \label{fermionH}

The spinor heat kernel on the hyperbolic space was evaluated in \cite{camp} as
 \bea
K_{\BH^{2n+1}}^{(1/2)}(t,x,y)&=&\hat U(x,y) \cosh\frac{\rho}{2}  \left({-1\over
2\pi R^2}\frac{\del} {\del \cosh \rho}\right)^n
\Big(\cosh\frac{\rho}{2}\Big)^{-1} { e^{-\frac{\rho^2 R^2}{4t}} \over (4\pi
t)^{1/2}},
 \label{diracheat}
 \\
K_{\BH^{2n+2}}^{(1/2)}(t,x,y)&=&\hat U(x,y) \cosh\frac{\rho}{2}  \left({-1\over
2\pi R^2}\frac{\del} {\del \cosh \rho}\right)^{n}
\Big(\cosh\frac{\rho}{2}\Big)^{-1} f_{\BH^2}^{(1/2)}(\rho, t),
 \label{diracheat2}
 \eea
where $x,y$ are two arbitrary points of the hyperbolic space; $n=0,1,2,...$;
$\rho$ is the geodesic distance between $x$ and $y$ in units of $R$; the matrix
$\hat U(x,y)$ is the parallel spinor propagator from  $x$ to $y$; and
 \be
f_{\BH^2}^{(1/2)}(\rho, t)={\sqrt{2}\,R \over (4\pi t)^{3/2} \cosh(\rho/2)}
\int_\rho^\infty {\tilde\rho \cosh {\tilde\rho \over 2}
e^{-{R^2\tilde\rho^2\over 4t} } \over \sqrt{\cosh\tilde\rho-\cosh\rho}} \,
d\tilde\rho ~.
  \label{2dheat}
 \ee

Alternatively, one can use $\zeta$-function method \reef{spartition3} to
evaluate the partition function $Z_\al$. The spinor $\zeta$-function on $\BH^d$
was computed in \cite{camp}. For odd $d\geq 3$
 \be
\zeta_{\BH^d}^{(1/2)}(z)={ 2^{[{d\over 2}]}R^{1-d}
m^{1-2z}\text{Vol}(\BH^{d})\over (4\pi)^{d/2} \Gamma(d/2)}
  \sum_{k=0}^{(d-1)/2} g_{k,d}^{(1/2)}\,
  (Rm)^{2k} B(k+1/2,z-k-1/2)
  \label{oddferzeta}
  ~,
 \ee
where $g_{k,d}^{(1/2)}$ are defined by
 \be
 \prod_{j=1/2}^{(d-2)/2}(x^2+j^2)=\sum_{k=0}^{(d-1)/2}g_{k,d}^{(1/2)}x^{2k}
 ~.
 \ee
On the other hand, for even $d\geq2$
 \begin{multline}
  \zeta_{\BH^d}^{(1/2)}(z)={ R^{2z-d}\text{Vol}(\BH^{d})\over (2\pi)^{d/2} \Gamma(d/2)}
  \sum_{k=0}^{(d-2)/2} h_{k,d}^{(1/2)}
  \Big[ (mR)^{2k+2-2z} B(k+1,z-k-1)
  \\
  +4 \int_0^\infty {x^{2k+1}\over (e^{2\pi x}-1)(x^2+m^2R^2)^{z}}dx\Big]
  \label{ferzeta}
  ~,
 \end{multline}
where $h_{0,2}^{(1/2)}=1$ and $h_{k,d}^{(1/2)}$ for even $d>2$ are defined by
 \be
 \prod_{j=1}^{(d-2)/2}(x^2+j^2)=\sum_{k=0}^{(d-2)/2}h_{k,d}^{(1/2)}x^{2k}
 ~.
 \ee

Clearly the $\zeta$-function approach \reef{spartition3} is the most efficient
way to determine the universal contributions in the \ren entropy \reef{renyi}.
However, as stressed previously, this method eliminates all of the power law
divergences, while the sharp cut-off in eq.~\reef{spartition} allows us to keep
track of the entire structure of the UV divergences appearing the \ren entropy.
Therefore, in what follows, we exploit both approaches to shed light on the
structure of the \ren entropy in the case of the spin-${1\over 2}$ field.

\subsubsection*{Odd dimensions}

It follows from  eq.~\reef{diracheat}  that for $d=2n+1$,
$K_{\BH^{d}}^{(1/2)}(x,x,t)$ takes the following general form
\begin{equation}
K_{\BH^{2n+1}}^{(1/2)}(x,x,t)=\frac{P_n^{(1/2)}(t/R^2)}{(4 \pi t)^{n+1/2} }
\  \mathbb{I}_{n+1},
\label{diracheat3}
\end{equation}
where $\mathbb{I}_{n+1}$ is a unit matrix of dimension $n+1$. This matrix is
the remnant of the spinor propagator $\hat U(x,y)$ in the limit of coincident
points. $P_{n}^{(1/2)}(x)$ is polynomial of degree $n$ with rational
coefficients
 \be
 P_{n}^{(1/2)}(x)=\sum_{j=0}^{n} a_{j,n}^{(1/2)}x^j ~,~ \text{for}\, n\geq 0\, .
 \ee
The first few of these polynomials are given by
 \bea
 P_0^{(1/2)}(x)&=& 1~,
 \nonumber \\
 P_1^{(1/2)}(x)&=&1+{1\over 2}x~,
  \nonumber \\
 P_2^{(1/2)}(x)&=&1+{5\over 3}x+{3\over 4}x^2~,
  \label{ferpoly} \\
 P_3^{(1/2)}(x)&=&1+{7\over 2}x+{259\over60}x^2+{15\over8}x^3~.
 \nonumber
 \eea
By definition $a_{0,n}^{(1/2)}\equiv 1$ and as we confirm below, in general,
 \be
a_{1,\frac{D-3}{2}}^{(1/2)}=\frac{(D-2)(D-3)}{12}\,.
 \label{lizz}
 \ee

Eq.~\reef{diracheat3} leads to the following expression for the \ren entropy,
 \bea
S_{\al}^{(1/2)}&=&{ 2^{n}(1+\al ) \over 24\,\al} \, \text{Vol}(\BH^{2n+1})
\int_{\delta^2}^{\infty} {dt\over t} \, \frac{P^{(1/2)}_{n}(t/R^2)}{(4 \pi
t)^{n+1/2} }\, e^{-m^2 t }~.
 \label{feroddren}
 \eea
The integral above is divergent in the vicinity of $t=0$. However, since
$P_n^{(1/2)}(x)$ is a polynomial of degree $n$ and the denominator contains a
half-integer power of $t$, all the divergences are simply power-law (and thus
non-universal). Using $a_{0,n}^{(1/2)}=1$, the leading contributions take the
form
 \be
S_{\al, div}^{(1/2)}= {(1+\al ) \over 12\al} {\sqrt{\pi}\,\cA_\Sigma \over
(D-2)\, (2\pi)^{D-1\over2}}   \Big( { 1\over \delta^{D-2}}-{D-2\over D-4}\Big(
m^2-{a_{1,{D-3\over 2}}^{(1/2)} \over R^2}  \Big){1\over \delta^{D-4}}
 +\ldots\Big)~,
  \label{farealaw}
 \ee
where $\cA_\Sigma=\text{Vol}(\BH^{D-2})$. Hence, the leading divergence
exhibits the expected `area law' behaviour. Continuing the process of expanding
the integrand in eq.~\reef{feroddren} near $t=0$, all power law divergences can
be evaluated for any fixed value of $D$. We can compare the above expression to
the results for a spherical waveguide for $D=3$ and 5 given eqs.~\reef{ferenS1}
and \reef{ferenS3}, respectively. Then up to expected substitution $R^2\to
-R^2$ in the curvature contributions, we see that the divergences in these two
different geometries agree.

To compute the finite part of the \ren entropy, one can use `dimensional
regularization' approach introduced in the previous section. In this scheme we
merely set $\delta=0$ and keep $n$ in eq.~\reef{feroddren} unspecified. As a
result, the power law divergences are eliminated and we find
 \bea
S_{\al,finite}^{(1/2)}&=& {(1+\al ) \over \al} {\cA_\Sigma  \over 48(2\pi)^{n
}\sqrt\pi}  \sum_{j=0}^n {a_{j,n}^{(1/2)} \over R^{2j} } \int_{0}^{\infty} dt
\,  t^{j-n-{3\over 2}} e^{-t \, m^2}
 \nonumber \\
&=&   {(1+\al ) \over \al} { \cA_\Sigma  \over 48(2\pi)^{n} \sqrt\pi } m^{2n+1}
\sum_{j=0}^n{a_{j,n}^{(1/2)} \over (m R)^{2j} } \, \Gamma(j-n-1/2)
 \label{frenyiodd} \\
&=&{ 1+\al \over 24\,\al} \,{  (-1)^{D-1\over 2}\pi\over (2\pi)^{D-2\over 2}
\Gamma(D/2)\sqrt 2} \cA_\Sigma \Big(m^{D-2}-a_{1,{D-3\over 2}}^{(1/2)}{D-2\over
2} \, {m^{D-4}\over R^2}+\cdots\Big)~,
 \nonumber
  \eea
where we have simplified the final expression with eq.~\reef{idnty}. Note that
we are expanding the final result in the limit $mR\gg1$. The leading
contribution in this limit precisely matches the expected area term
\reef{franks} with the pre-factor given by eq.~\reef{coeff2}.

The sub-leading term above also agrees with eqs.~\reef{ferenS1univ} and
\reef{ferenS3univ} for $D=3$ and 5, respectively, if we replace $R^2\to -R^2$
and substitute $a_{1,0}^{(1/2)}=0$ and $a_{1,1}^{(1/2)}=1/2$ using
eq.~\reef{ferpoly}. We might remark that, in fact, the $\zeta$-function method
can also be applied here to derive the same result. In particular, this
calculation confirms the result $a_{1,{D-3\over 2}}^{(1/2)}$ given in
eq.~\reef{lizz}. Following our discussion for the scalar fields, we can use
this general result to identify the form of this subleading term for arbitrary
$D$. We can then re-express this contribution in the covariant form:
 \be
S_\mt{univ}=\frac{D-2}{24}\,\gamma_{D,fermion}\,\int_\Sigma \!d^{D-2}\!\sigma\,
\sqrt{h}\ \cR(h)\ m^{D-4}\,.
 \label{firstR12}
 \ee
Again $\cR(h)$ is the Ricci scalar of the metric induced on the entangling
surface while the coefficient $\gamma_{D,fermion}$ is given in
eq.~\reef{coeff2}. This expression is only applicable for odd $D\ge5$.

\subsubsection*{Even dimensions}

The computation of the \ren entropy for even dimensions is, of course, similar
to that for odd dimensions. However, a systematic evaluation of all the
divergences using the simple regularization of the heat kernel
\reef{spartition} requires more effort for even $d$. The extra computational
complications originate from the fact that heat kernel of the Dirac operator on
$\BH^2$ cannot be expressed in terms of elementary functions. Instead we have
the integral representation in eq.~\reef{2dheat}.

To demonstrate the procedure, we start from the special case $d=2$ and evaluate
all the divergences in this case. The extension of this computation to higher
even $d$ presents no conceptual difficulties. Therefore rather than pursuing
this approach \reef{spartition} to construct a  voluminous general expression
which contains all of the (nonuniversal) power law divergences for general even
$d$, we shift our focus to only evaluating the universal logarithmic
contributions. In this case, the $\zeta$-function approach \reef{spartition3}
provides the most efficient method to produce a general result.
\newline
\\
{\it $d=2$ ($D=4$):}
\\
\newline
\\
In this case we need  eq.~\reef{diracheat2} with $n=0$. In particular then, the
limit of coincident points gives
 \be
 K_{\BH^2}^{(1/2)}(t,x,x)= f_{\BH^2}^{(1/2)}(0,t)\, \mathbb{I}_2 ~.
 \ee
Again the two-by-two identity matrix $\mathbb{I}_2$ corresponds the coincident
point limit of the spinor propagator $\hat U(x,y)$ on $\BH^2$. Given the
expression in eq.~\reef{2dheat}, we evaluate $f_{\BH^2}^{(1/2)}(0,t)$ by
expanding for small $t$
 \bea
f_{\BH^2}^{(1/2)}(0, t)&=&{ 1 \over (4\pi)^{3\over 2}  R \sqrt{t}}\int_0^\infty
 x \, e^{-x^2/4}\coth \Big({\sqrt{t} \, x \over 2 R}\Big)  \, dx
  \nonumber \\
  &=&{ 1 \over 4\pi t } \Big( 1 + {t\over 6 R^2} + \ldots\Big)
  ~.
 \eea
Here we only explicitly show the terms which contribute to the divergences in
the \ren entropy and the ellipsis denotes terms which only make finite
contributions. Proceeding as usual with this result, we find
 \bea
S_{\al}^{(1/2)}&=& {(1+\al) \over \al}  {\text{Vol}(\BH^{2}) \over 48 \pi }
\int_{\delta^2}^{\infty} {dt\over t^2} \, e^{-t \, m^2} \Big( 1 + {t\over 6
R^2} + \ldots\Big)~,
 \nonumber \\
&=&  {1+\al\over 48\pi\al}\,\cA_\Sigma \,\( {1 \over \, \delta^{2} } +\Big( 2
m^2-{1 \over 3 R^2} \Big) \log(m\delta)+...\)
  \label{ferren4d}
 \eea
where $\cA_\Sigma=\text{Vol}(\BH^2)$. Of course, the leading divergence above
corresponds to the usual `area law' term. For $\al=1$, the second term
precisely matches eq.~\reef{franks} with pre-factor given by eq.~\reef{coeff2}
with $D=4$. Finally, eq.~\reef{ferren4d} agrees with the analogous result
\reef{ferren4dS} for a spherical waveguide after replacing  $R^2\to -R^2$.
\newline
\\
{\it General even $d\geq 4$}
\newline
\\
Here we use the spinor $\zeta$-function \reef{ferzeta} to compute the partition
fucntion \reef{spartition3} and then the \ren entropy \reef{renyi}. Evaluating
eq.~\reef{ferzeta}  at $z=0$ and substituting the result into
eq.~\reef{spartition3} leads to the following universal contribution to the
\ren entropy
 \begin{multline}
S_{\al,univ}^{(1/2)}=-{(1+\al) \over \al} { \text{Vol}(\BH^{d})\over
12(2\pi)^{d/2} \Gamma(d/2)R^{d}}
\\
\times\sum_{k=0}^{(d-2)/2}
h_{k,d}^{(1/2)}\[  4 \int_0^\infty {x^{2k+1}\over e^{2\pi x}-1}dx+
{(-R^2m^2)^{k+1} \over k+1}\] \log(m\delta)~.
 \end{multline}
Substituting $d=2$, we recover the universal part appearing in
eq.~\reef{ferren4d} above. Substituting in other explicit values of $d$, one
can generate universal contributions for higher even dimensions. Thus, for
instance, we obtain with $d=4\ (D=6)$
 \be
S_{\al,univ}^{(1/2)}(d=4)=-{(1+\al) \over 96\pi^2\al} \text{Vol}(\BH^{4}) \(
m^4- 2{m^2\over R^{2}}+{11\over 30}{1\over R^{4}}\)\log(m\delta)
 ~.
 \label{6Dferm}
 \ee

We can also expand the above expression in the limit $mR\gg 1$ to determine a
general expression for the leading terms:
 \be
S_{\al,univ}^{(1/2)}={(1+\al) \over 12\al} \frac{(-)^{D/2}
\cA_\Sigma}{(2\pi)^{\frac{D-2}{2}}\,\Gamma(D/2)}\(m^{D-2}-{(D-2)^2(D-3)\over
24} \, {m^{D-4}\over R^2}+\cdots\)\log(m\delta)~,
 \ee
where $\cA_\Sigma=\text{Vol}(\BH^{D-2})$. The leading term has the expected
form \reef{franks} with the pre-factor given \reef{coeff2}. The next-to-leading
term reveals a new universal curvature contribution \reef{univEE}. Let us turn
to the entanglement entropy by setting $\alpha=1$ and then write this
subleading contributions in a covariant form as
 \be
S_\mt{univ}=\frac{D-2}{24}\,\gamma_{D,fermion}\,\int_\Sigma \!d^{D-2}\!\sigma\,
\sqrt{h}\ \cR(h)\ m^{D-4}\,\log(m\delta)\,.
 \label{firstRx12}
 \ee
Again $\cR(h)$ is the Ricci scalar of the metric induced on the entangling
surface and the coefficient $\gamma_{D,fermion}$ is given in eq.~\reef{coeff2}.
Of course, this contribution only appears for even $D\ge4$.

\end{document}